\documentclass{aa}  

\usepackage[normalem]{ulem}
\usepackage{graphicx}
\usepackage{float}         
\usepackage{natbib}
\bibpunct{(}{)}{;}{a}{}{,} 
\usepackage[varg]{txfonts}
\usepackage{xcolor}

\usepackage{verbatim}
\usepackage{subcaption}
\newcommand{\abund}[2]{\ensuremath{[\mathrm{#1}/\mathrm{#2}]}}
\newcommand{\teff}{\ensuremath{T_\mathrm{eff}}}
\newcommand{\logg}{\ensuremath{\log\,g}}
\newcommand{\metal}{\abund{Fe}{H}}
\begin{document} 

 \title{J-PLUS: Searching for very metal-poor star candidates using the SPEEM pipeline}

   \author{Carlos Andrés Galarza \inst{1},
          Simone Daflon\inst{1},
          Vinicius M. Placco\inst{2},
          Carlos Allende Prieto\inst{3,4},
          Marcelo Borges Fernandes  \inst{1},
          Haibo Yuan\inst{5},
          Carlos López-Sanjuan\inst{6},
          Young Sun Lee\inst{7},
          Enrique Solano\inst{8}, F. Jim\'enez-Esteban\inst{8},
          David Sobral\inst{9},
          Alvaro Alvarez Candal\inst{1, 17},
          Claudio B. Pereira\inst{1}, 
          Stavros Akras\inst{10}, 
          Eduardo Martín\inst{3, 8},
          Yolanda Jiménez Teja\inst{16},
          Javier Cenarro\inst{6},
          David Cristóbal-Hornillos\inst{6},
          Carlos Hernández-Monteagudo\inst{6},
          Antonio Marín-Franch\inst{6},
          Mariano Moles\inst{6},
          Jesús Varela\inst{6},
          Héctor Vázquez Ramió\inst{6},
          Jailson Alcaniz\inst{1},
          Renato Dupke\inst{1, 13, 14, 15},
          Alessandro Ederoclite\inst{11},
          Laerte Sodré Jr.\inst{11},
          \and
          Raul E. Angulo\inst{12}
          }

   \institute{Observat\'{o}rio Nacional - MCTI (ON), Rua Gal. Jos\'{e} Cristino 77, S\~{a}o Crist\'{o}v\~{a}o, 20921-400, Rio de Janeiro, Brazil\\
              \email{carlosgalarza@on.br}
         \and
             NSF’s NOIRLab, 950 N. Cherry Ave., Tucson, AZ 85719, USA
        \and
            Instituto de Astrofísica de Canarias, Vía Láctea, 38205 La Laguna, Tenerife, Spain
        \and
            Universidad de La Laguna, Departamento de Astrofísica, 38206 La Laguna, Tenerife, Spain
        \and    
            Department of Astronomy, Beijing Normal University, Beijing 100875, People’s Republic of China
             \and
            Centro  de  Estudios  de  Física  del  Cosmos  de  Aragón  (CEFCA),Unidad Asociada al CSIC, Plaza San Juan 1, E-44001, Teruel, Spain
        \and
            Department of Astronomy and Space Science, Chungnam National University, Daejeon 34134, South Korea
        \and
        Departamento de Astrof\'{\i}sica, Centro de Astrobiolog\'{\i}a (CSIC-INTA), ESAC Campus, Camino Bajo del Castillo s/n, E-28692 Villanueva de la Ca\~nada, Madrid, Spain
        \and
        Department of Physics, Lancaster University, Lancaster, LA1 4YB, UK
        \and
        Institute for Astronomy, Astrophysics, Space Applications and Remote Sensing, National Observatory of Athens, GR 15236 Penteli, Greece
        \and
            Universidade  de  São  Paulo,  Instituto  de  Astronomia,  Geofísica e Ciências Atmosféricas, R. do Matão 1226, 05508-090, São Paulo,Brazil
        \and
            Ikerbasque, Basque Foundation for Science, E-48013 Bilbao, Spain
        \and 
        Department of Astronomy, University of Michigan, 930 Dennison Bldg., Ann Arbor, MI 48109-1090, USA
        \and
        Eureka Scientific Inc., 2452 Delmer St. Suite 100, Oakland, CA 94602, USA
        \and
        Department of Physics and Astronomy, University of Alabama, Box 870324, Tuscaloosa, AL 35487, USA
        \and
        Instituto de Astrof\'isica de Andaluc\'ia--CSIC, Glorieta de la Astronom\'ia s/n, E--18008 Granada, Spain
        \and
        Universidad de Alicante, Carr. de San Vicente del Raspeig, s/n, 03690 San Vicente del Raspeig, Alicante, Spain
             }

   \date{Received XXXX ?, 2021; accepted ? ?, ?}

 
  \abstract
   {We explore the stellar content of the Javalambre Photometric Local Universe Survey (J-PLUS) Data Release 2 and show its potential to identify low-metallicity stars using the Stellar Parameters Estimation based on Ensemble Methods (SPEEM) pipeline.}
   {SPEEM is a tool  to provide determinations of  atmospheric parameters for stars and separate stellar sources from quasars, using the unique J-PLUS photometric system. The adoption of adequate selection criteria allows the identification of metal-poor star candidates suitable for spectroscopic follow-up.}
   {SPEEM consists of a series of machine learning models which uses a training sample observed by both J-PLUS and the SEGUE spectroscopic survey. The training sample has temperatures \teff\ between 4\,800 K and 9\,000 K; \logg\ between 1.0 and 4.5, and $-3.1<$ \metal $<+0.5$. 
   The performance of the pipeline has been tested with a sample of stars observed by the LAMOST survey within the same parameter range.}
   {The average differences between the parameters of a sample of stars observed with SEGUE and J-PLUS, which were obtained with the SEGUE Stellar Parameter Pipeline and SPEEM, respectively, are $\Delta$\teff$\sim$41~K, $\Delta$\logg$\sim$0.11~dex, and $\Delta$\metal$\sim$0.09~dex. A sample of 177 stars have been identified as new candidates with \metal $<-$2.5 and 11 of them have been observed with the ISIS spectrograph at the William Herschel Telescope. The spectroscopic analysis confirms that 64\% of stars have \metal$<-2.5$, including one new star with \metal$<-$3.0.}
   {SPEEM in combination with the J-PLUS filter system has shown the potential to estimate the stellar atmospheric parameters (\teff, \logg, and \metal). The spectroscopic validation of the candidates shows that SPEEM yields a success rate of 64\% on the identification of very metal-poor star candidates with \metal $<-$2.5.}
   

   \titlerunning{Searching for VMPs using SPEEM}
   \authorrunning{Galarza et al.}
   \maketitle
%

\section{Introduction}

The electromagnetic radiation emitted by a stellar source and collected by a telescope  allows the determination of multiple physical quantities, such as  the effective temperature (\teff), surface gravity (\logg), and metallicity  \metal\footnote{[A/B] = $\log(N_A/{}N_B)_{\star} - \log(N_A/{}N_B) _{\odot}$, where $N$ is the number density of chemical elements $A$ and $B$ in the star ($\star$) and the Sun ($\odot$).}. Thus, the analysis of stellar spectra permits the determination of stellar parameters and  abundances for many chemical species. However, for stars without spectra accessible, narrow-band photometry offers an alternative approach to determining abundances for selected elements, such as carbon \citep{whitten2021}.

One of the main questions that astronomical observations can help to answer is how the formation of our Galaxy took place and how it evolved to its current state. 
For that purpose, the identification of Very Metal-Poor (VMP, \metal$<-2.0$) and Extremely Metal-Poor stars (EMP, \metal$<-3.0$), according to \citet{beers2005}, represents a critical step for understanding the origin of the Milky Way. However, according to the SAGA\footnote{http://sagadatabase.jp/} database as updated on April 2021, the lower end of the halo metallicity distribution function contains $\sim$550 EMPs, being  a few dozen stars with \metal\ $<-4$ confirmed spectroscopically.
The identification of new objects of this type is attainable by using different approaches based on photometry and spectroscopy. In contrast to spectroscopy, photometry offers the advantage of higher signal-to-noise ratios for a given exposure time and the simultaneous collection of data from a large number of sources  but, in turn, photometric data  is not able to provide characterization of  individual spectral features.

Over the last two decades, wide-field broad-band photometric surveys, such as the Sloan Digital Sky Survey \citep[SDSS;][]{york}, and the Panoramic Survey Telescope and Rapid Response System \citep[Pan-STARSS;][]{2016arXivPanStarss} based on ground-based telescopes, have proven to be an extremely successful alternative to challenging spectroscopic surveys when it comes to studying large amounts of astronomical objects. 

On the other hand, good examples of spectroscopic surveys employed to search for very metal-poor stars are:
The Apache Point Observatory Galactic Evolution Experiment \citep[APOGEE;][]{majewski2016};
The Sloan Extension for Galactic Understanding and Exploration \citep[SEGUE;][]{yanny2009segue};
The Large Sky Area Multi-Object Fiber Spectroscopic Telescope \citep[LAMOST;][]{LAMOST_cui}; 
and the upcoming WHT Enhanced Area Velocity Explorer \citep[WEAVE;][]{dalton2012} and The 4-meter Multi-Object Spectroscopic Telescope \citep[4MOST;][]{4most}.

More recently, many exciting projects are entering the scene, as for example 
The Javalambre Photometric Local Universe Survey \citep[J-PLUS;][]{JPLUS} and the Javalambre Physics of the Accelerating Universe Astrophysical Survey \citep[J-PAS;][]{JPAS} 
covering several thousands of deg$^2$ of the Northern sky. The unique systems of 12 and 60 filters for J-PLUS and J-PAS, respectively, allow accurate estimations of stellar parameters, providing information on some critical spectral features framed by some of the  narrow-band filters. In the southern sky, the Southern Photometric Local Universe Survey \citep[S-PLUS;][]{splus,splus_dr2} and the SkyMapper Southern Sky Survey \citep[SkyMapper;][]{skymapper}, also rely on a combination of narrow and broad-band photometry to obtain data.
Another interesting ongoing project focused on the northern hemisphere is the Pristine survey \citep{pristine1}, which uses an exclusive narrow-band filter centered on \ion{Ca}{ii} H\&K lines combined with the Sloan broad-band filters. 

Photometric and spectroscopic data complement each other, meaning that the best possible scenario for astronomical research is to combine both types of data analyses. Different photometric calibrations present reliable results to estimate stellar parameters but with a restricted range of applicability. For instance, \citet{ivezic2008} found that $(g-r)$ color from SDSS provides an accurate estimation of \teff, while \metal\ can be obtained through a polynomial fitting using $(u-g)$ with some restrictions on $(g-r)$, due to a rapid saturation of the blue band, and losing precision for stars with \metal$<-2.0$. Another independent methodology to infer \teff\ is the Infrared Flux Method \citep[IRFM;][]{casagrande2006}, which can be used to build relantionships between  the flux measured in the infrared and BVJHK photometry. These calibrations are valid for \teff =   4\,000 -- 8\,000 K, corresponding to stars with spectral types F, G, and K, with \metal\ varying from  $-5$ to 0.4.  
In terms of spectroscopic observations,~Lee et al.~(\citeyear{lee2008seguea,lee2008segueb}) presented the SEGUE Stellar Parameter Pipeline (SSPP), which estimates stellar parameters using both theoretical and empiric calibrations of medium resolution stellar spectra ($R\sim 1\,800$) in combination with the implementation of neural networks and validations based on spectral libraries such as ELODIE \citep{prugniel2001database,moultaka2004elodie} and MILES \citep{MILES}, and high-resolution spectra for additional validation (Allende-Prieto et al. \citeyear{prieto2008segue}). \citet{lasp} and \citet{xiang2015lamost} also presented similar pipelines (LASP and LSP3) to estimate parameters from spectra produced by the LAMOST survey.
   
In the upcoming years, the datasets that will be produced by different extensive surveys like the Large Synoptic Survey Telescope (LSST) and the progressive data releases of the Gaia mission \citep{prusti2016gaia,gaia2018gaia} will be massive so that new effective strategies to deal with the data will be needed. In that sense, machine learning algorithms or any other statistical tools may be useful to analyze data and make reasonably quick predictions. Machine learning methods such as artificial neural networks (ANN) are part of the modern approach for various astronomical applications. For instance, \citet{singh1998stellar} presented a model to classify stellar spectra; more recently, \citet{whitten2019j} used ANNs to estimate \teff\ and \metal\ for J-PLUS data in order to search for low-metallicity stars. Other models based on other algorithms such as Random Forest \citep[hereafter RF;][]{Breiman2001} and Extreme Gradient Boosting \citep[hereafter XGB;][]{chen2015xgboost,chen2016xgboost} show promising results of morphological classifications and estimation of physical parameters. \citet{Miller_2015} presented a RF model capable of inferring \teff, \logg\, and \metal\ based on SDSS de-reddened colors. More recently, \citet{bai2018machine} built a RF model that performed a Star-Galaxy-QSO classification and calculated the \teff\ for stars using data from SDSS and LAMOST, while \citet{chao2019study} applied a model based on the XGB algorithm able to classify stars and galaxies in the SDSS improving the results specially on the faint light sources. The search for VMPs candidates from wide-angle photometric surveys like J-PLUS, complemented with spectroscopic follow-up, can benefit enormously from machine learning methods. 

This paper presents the Stellar Parameters Estimation based on the Ensemble Methods pipeline (hereafter  SPEEM) that explores the benefits of RF and XGB algorithms and how it offers an excellent alternative to analyze J-PLUS data by estimating three main physical parameters of stellar sources: \teff, \logg, and \metal. 
Section 2 presents the databases employed to produce the training and validation samples required in the development of SPEEM. 
Section 3 covers the details regarding the pipeline architecture and the cleaning process centered on removing contaminants, such as extragalactic point-like sources (QSOs) and white dwarfs (WD). In Section 4 we explain the training and validation process, pointing out the features selected as reliable indicators of the stellar parameters. Section 5 describes how SPEEM helped to select interesting very metal-poor stars suitable for spectroscopic follow-up observations that resulted in 11 metal-poor stars candidates, being 10 newly discovered stars and  one of them being an extremely metal-poor star. Finally, Section 6 summarizes the overall results obtained and future applications and improvements to extend estimation of metallicities to the \metal\ $< -3.0$ regime.
   
 \section{Datasets}

This section describes the main datasets used to study the correlations between J-PLUS colors and the parameters estimated independently by other surveys.

 \subsection{J-PLUS photometric data}
   
J-PLUS\footnote{\url{www.j-plus.es}} is being conducted from the Observatorio Astrof\'{\i}sico de Javalambre (OAJ, Teruel, Spain; \citealt{oaj}) using the 83\,cm Javalambre Auxiliary Survey Telescope (JAST80) and T80Cam, a panoramic camera of 9.2k $\times$ 9.2k pixels that provides a $2\deg^2$ field of view with a pixel scale of 0.55 arcsec pix$^{-1}$ \citep{t80cam}. The J-PLUS filter system is composed of twelve passbands (Table 1). The J-PLUS observational strategy, image reduction, and main scientific goals are presented in \citet{JPLUS}.

The J-PLUS second data release (DR2) comprises $1\,088$ pointings ($2\,176$ deg$^2$) observed and reduced in all survey bands \citep{jplus_dr2}. The photometric calibration was performed using both the metallicity-dependent stellar locus and the white dwarf locus (López-Sanjuan et al. 2019a, 2021). The limiting magnitudes (5$\sigma$, 3 arcsec aperture, AB system) of the DR2 are $\sim 21$ mag in $g$ and $r$ passbands, and $\sim 20$ mag in the other nine bands. The median point spread function (PSF) full width at half maximum (FWHM) in the DR2 $r$-band images is 1.1 arcsec. Source detection was done in the $r$ band using \texttt{SExtractor} \citep{sextractor}, and the flux measurement in the twelve J-PLUS bands was performed at the position of the detected sources using the aperture defined in the $r$-band image. The DR2 is publicly available at the J-PLUS website\footnote{\url{www.j-plus.es/datareleases/data_release_dr2}}.

    \begin{table}[H]
    \centering
    \caption[ ]{\label{table:jplusfilters}
    {J-PLUS Filter System.}}
    \begin{tabular}{cccc}\hline
    \hline
             & Central &  &  \\     
     Filter  &  Wavelength & FWHM & Main  \\
             &  (\AA) & (\AA) &  Features\\
    \hline
    \textit{u}     &	3\,485 & 508    &     \\		
	    J0378      &	3\,785 & 168    & \ion{O}{ii}\\		
    	J0395      &    3\,950 & 100    & \ion{Ca}{ii} H\&K \\	
	    J0410      &	4\,100 & 200    & H$\delta$ \\
	    J0430      &	4\,300 & 200    & G-band \\		
    \textit{g}     &	4\,803 & 1\,409 &   \\		
	    J0515      &	5\,150 & 200    & Mg Ib Triplet\\	
	\textit{r}     &	6\,254 & 1\,388 &   \\		
	    J0660      &	6\,600 & 145    & H$\alpha$\\	
	\textit{i}     &	7\,668 & 1\,535 &   \\		
	    J0861      &	8\,610 & 400    & \ion{Ca}{ii} Triplet\\
	\textit{z}     &	9\,114 & 1\,409 &   \\		
    \hline
    \end{tabular}
   \end{table}

In order to select the sample for our study and ensure high-quality  measurements  in  each  of  the  twelve  filters,  we  chose  6 arcsec aperture magnitudes plus aperture correction and a proper configuration in the flags parameters, such as \texttt{MASKS\_FLAGS} = 0 (indicating the target is not inside a Mask), \texttt{FLAGS} = 0 (no Sextractor flags detected), and \texttt{NORM\_WMAP\_VAL} > 0.8 (indicating adequate exposure). We corrected by interstellar reddening using the extinction correction vector $A_x$ reported in the J-PLUS DR2 database.  Considering the morphological classification star/galaxy based on PDF analysis proposed by~\citet{jplus_morph_class}, we selected all targets with the probability of being a star $p_{\rm star}>0.9$. In addition, we selected stars brighter than \textit{g} $<18$, resulting in a sample of 746\,531 objects (hereafter referred to as Gold sample).

\subsection{The Sloan Extension for Galactic Understanding and Exploration - SEGUE}
   
SEGUE~\citep{yanny2009segue} is part of the second phase of the Sloan survey  (SDSS-II), and consists of a set of  250\,000 medium resolution spectra of stars observed within 3\,500  deg$^2$ of the northern sky, excluding regions at low galactic latitude ($|l<30^{o}|$). The wavelength coverage is between 3\,900\AA\, and 9\,000\AA\, at $R\sim1\,800$, with most of the  observed stars presenting spectral types from A to M. SEGUE has produced several important  results, including the identification of stars at  $\metal <-3.0$ \citep{aoki2012high,placco2015metal}, the study of the structure of the Galactic thick disk and halo \citep{jong2010,lee2017chemical,lee2019chemical,kim2019dependence,kim2021evidence}, the analysis of \abund{\alpha}{Fe} ratio of G-dwarfs from the Galactic disk \citep{lee2011}, and the study of carbon-to-iron ratio of dwarfs, main sequence turnoff stars, and giants in the Milky Way \citep{lee2013carbon}.
   
The SEGUE Stellar Parameter Pipeline (SSPP) estimates the stellar parameters and abundances of selected elements from SEGUE spectra. It employs a combination of  multiple techniques such as spectral fitting (Allende Prieto et al.~\citeyear{Allende_Prieto_2006}) and minimization of $\chi^{2}$ within grids of model atmospheres, analysis of \ion{Ca}{ii} K lines, autocorrelation functions~\citep{beers1999estimation}, calibrations of \ion{Ca}{ii} triplet~\citep{cenarro2001empirical,cenarro2001_b}, implementation of artificial neural networks trained both on  observed ~\citep{fiorentin2007estimation} and synthetic spectra, and semiempirical predictions using \textit{g}$-$\textit{r} color (valid only for \teff). 
   
\subsection{The LAMOST Experiment for Galactic Understanding and Exploration - LEGUE}
   
LEGUE is a survey of the Galactic structure within the LAMOST project~\citep{deng2012lamost}. The multi-object spectrograph allows to obtain spectra with R$\sim 1\,800$ for 4\,000 targets simultaneously.  LEGUE will provide a sample of 5 million stars  with limited sky coverage. One of the LEGUE primary science goals is the search for EMPs. The LAMOST Stellar Parameter Pipeline \citep[LASP;][]{lasp} has been used to estimate stellar parameters from LAMOST data. LASP used the Universit\'e de Lyon Spectroscopic Analysis Software \citep[ULYSS;][]{koleva2009ulyss,wu2011automatic} jointly with the ELODIE library. This pipeline has been successfully tested and applied to estimate radial velocities and stellar parameters to LAMOST data.  Value Added Catalogs are available with official data releases~\citep{luo2015first}. The LAMOST Stellar Parameter Pipeline at Peking University LSP3~\citep{xiang2015lamost} is an alternative pipeline functional to calculate stellar parameters, also tested with LAMOST data. LSP3 uses the MILES library in addition to weighted average parameters that best match the spectra templates and values yielded by $\chi^{2}$ minimization. 
   
 \subsection{Sample Cross-matches}
 
  The J-PLUS Gold sample was cross-matched with the surveys  SEGUE and LAMOST DR5 to retrieve information about physical parameters and morphological classification for the targets in common. That process produced the following datasets:
   
   \begin{itemize}
  \item J-PLUS$\times$SEGUE, with  6\,794 targets in
  common, contains information on stellar parameters (\teff, \logg, \metal) estimated by the SSPP pipeline.
   
   \item J-PLUS$\times$LAMOST, with 99\,184 targets with signal-to-noise $>10$. 
   The analysis presented in this paper is anchored on estimates of atmospheric parameters obtained with the n-SSPP pipeline \citep{Beers_2014,beers2017}, an extension of  the SEGUE pipeline, using a grid of spectra with $\chi^2$ minimization to choose the best fit to the observed spectrum, providing \teff, \logg\ and \metal.  
   
   \end{itemize}
 
  \section{Model Training}
  
   SPEEM is a set of machine learning models developed from a supervised training approach, starting with a subsample of objects with previously known parameters such as classification star/QSO,  spectral type, \teff, \logg, and \metal. The training
   process explores statistical relationships between the input parameters and measured features, such as magnitudes and color indices. An extended discussion about supervised machine learning techniques is available in \citet{kotsiantis2007supervised}. These relationships allow us to predict the stellar parameters of the Gold sample stars.
  
  \subsection{Decision Trees, Random Forest and Extreme Gradient Boosting}
  
  Decision trees are part of the most useful and intuituive tools used to deploy machine learning models for classification and regression tasks, usually referred to as Classification and  Regression Trees   \citep[CARTs;][]{breiman1984classification}. CARTs are easy to train and interpret.  However they tend to overfit in some cases, for instance, on application to datasets with unbalanced classes
  producing lower-precision estimations \citep{strobl2009introduction}.
  
  A CART structure is hierarchical (top-down nodes) and composed of a subset of the training sample derived recursively. This splitting process continues until the subset at a node is no longer statistically meaningful regarding the relationship between the
  input and the target variables. In order to build a classification tree, it is necessary to define an information gain parameter to be optimized by the algorithm to get accurate estimations. This parameter usually comes from the Gini impurity or Shanon entropy coefficients,  expressed by equations (1) and (2), respectively, 
 
     $$G(X_{n}) = \sum_{k} p_{nk}(1-p_{nk}), \eqno(1)$$
     $$H(X_{n}) = \sum_{k} p_{nk}\log(p_{nk}), \eqno(2)$$
 
  \noindent where $X_{n}$ is the training data in node $n$ and $p_{nk}$ is  the proportion of class $k$ observed values in node $n$. For regression purposes, the criteria to be minimized is usually the mean squared error (MSE), expressed by

 $$\textrm{MSE}= \dfrac{1}{n} \sum_{i=0}^{n-1}\Delta^{2}, \eqno(3)$$
  
  \noindent where $\Delta=X_{i}^{Target}-X_{i}^{Model}$ is the difference between reference value $X_{i}^{Target}$ and  the  value estimated by machine learning $X_{i}^{Model}$, and $n$ is the sample size. 
  
Random Forest (RF) represents an evolution of decision trees, since it combines an arbitrary number of trees, each one of them fed with a random subsample of the input data and fitted with random subsets of features  to avoid bias selection and prevent overfitting, offering more robust estimations. On the other hand, the Extreme Gradient Boosting (XGB) can outperform RF models by allowing the possibility of penalizing the trees towards learning the more difficult data, using a combination of parallel and sequential computing (i.e latter trees or forest learn from the errors made by previous ones).

A critical concern on the use of decision trees relies on the distribution of parameters of the studied sample. For example, the metallicity distribution obtained in the Pristine Survey for the inner Galactic halo peaks at \metal\  $=-1.6$ \citep{pristine}
 and  the number of stars decreases towards lower \metal\ values with a slope of $\Delta (\log {\textrm N} )/\Delta $\metal=1.0$\pm$0.1. Since we are primarily interested in stars with \metal $< -2.2$, which are more scarce than the bulk of the distribution, the training datasets for different  metallicity bins are unbalanced. Adopting decision trees on such unbalanced datasets may be a disadvantage but RF is likely a reasonable alternative. Despite RFs performing well in estimating the parameters, we decided to apply the XGB algorithm to the regression process of the photometric parameters. That resulted in a significant improvement in dealing with unbalanced dataset, as it is the case for the metallicity distribution of the Galactic Halo. The following sections discuss the precision and the accuracy obtained by SPEEM.
  
  \begin{figure}[H]
    \centering
  \includegraphics[scale=0.4]{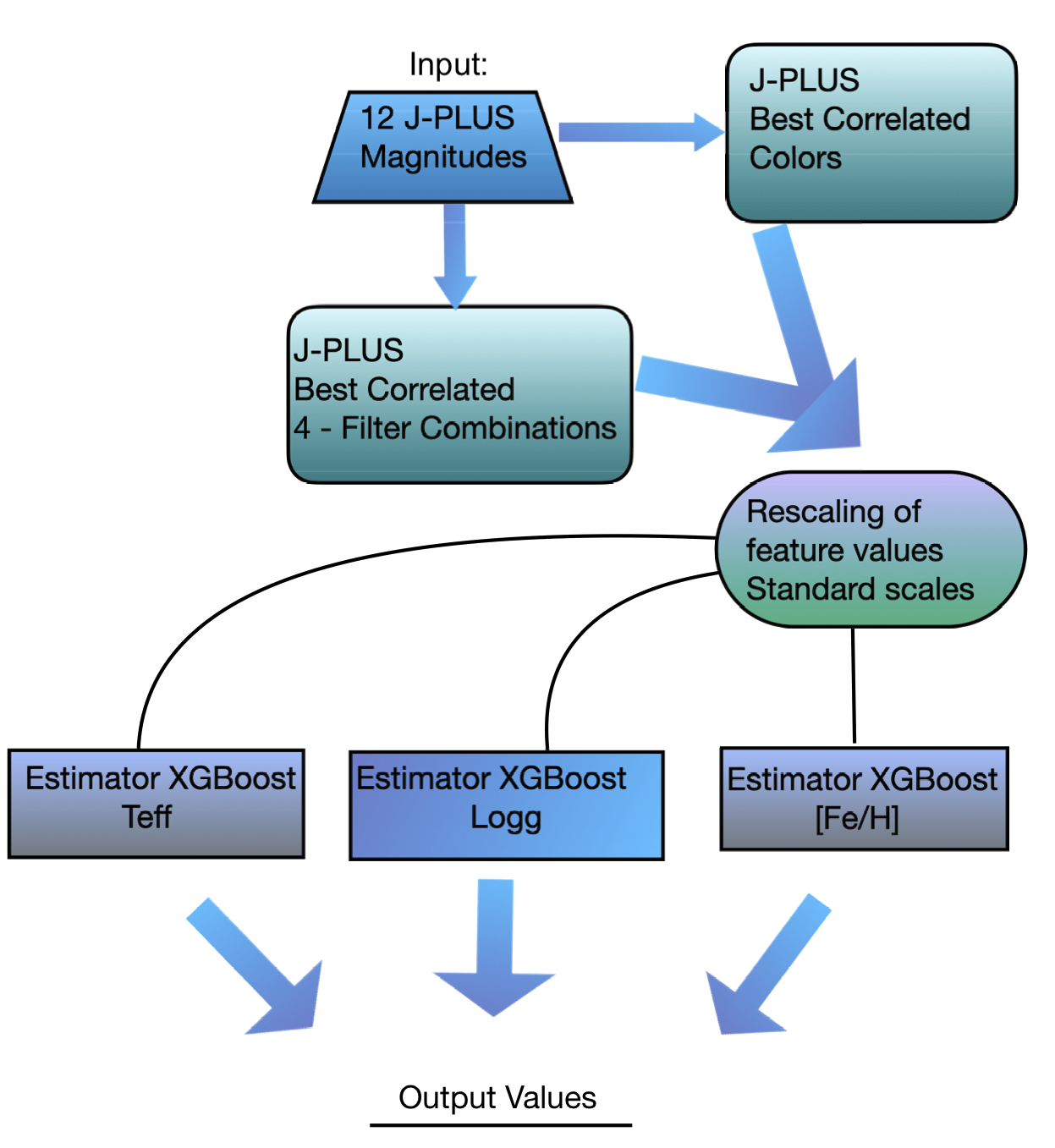}
    \caption{SPEEM strucuture showing the three parts of the pipeline: selection of  features, normalization and the multi-output regressor.}
    \label{pipeline_structure}
    \vspace*{-0.3cm}
  \end{figure}   
  
  \subsection{SPEEM Architecture}
  
  \begin{figure*}[ht!]
  \centering
  \begin{subfigure}[]{0.54\textwidth}
    \centering
    \includegraphics[width=\textwidth]{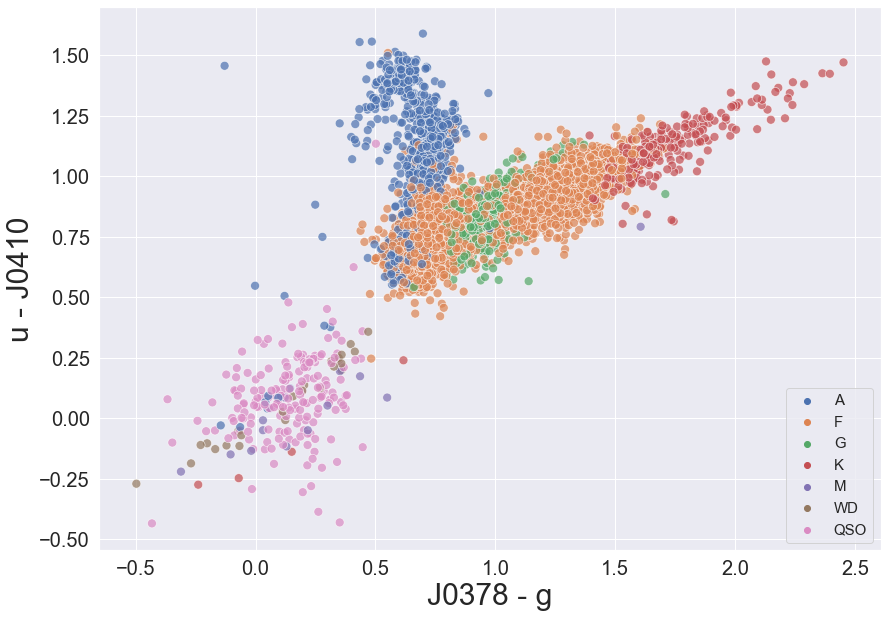}
  \end{subfigure}
  \hfill
  \begin{subfigure}[]{0.42\textwidth}
    \centering
    \includegraphics[width=\textwidth]{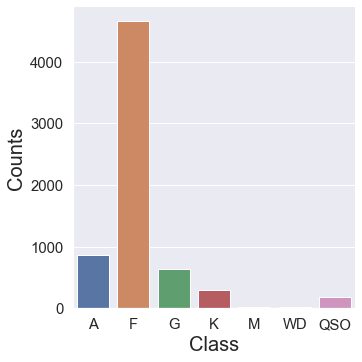}
  \end{subfigure}
  \hfill
  \caption{{\it Left}: QSO-Star separation based on the the J-PLUS colors  \textit{u}$-$J0410 and J0378$-$\textit{g} for a sample of 6\,794 stars. A threshold line at \textit{u}$-$J0410=0.5 can be used to separate QSOs (pink filled circles) and stars (color-coded according to the spectral type). 
  {\it Right}: Distribution of classes (stellar spectral types and QSO) assigned by SEGUE  for the studied sample. 
  }
  \label{fig:star_QSO}
  \end{figure*}
  
  All the algorithms used in the development of SPEEM are available on the \texttt{Scikit-learn} package for the \texttt{Python} programming language  \citep{scikit-learn}.
  
The  SPEEM architecture, as shown in  Figure~\ref{pipeline_structure}, consists on three different parts: the  features selection, rescaling of the features (normalization) and the multi-output regressor, that gives the final output. 

  The first part of the pipeline, the features selection, is activated by applying the Feature Union module, which creates a customized set of features composed of photometric magnitudes, colors and combination of colors. These features are used by the pipeline as input to make the spectral type classification and the estimation of  stellar parameters. Then, the Standard Scaler function transforms the distribution of the parameters into normal distributions centered around $0$ with a standard deviation of $1$, to avoid any biases. The model considers the contribution of each variable without previous statistical assumptions due to the difference in scale values. Hence, all the scaled features are passed through the third part of the pipeline, the multi-output regressor, which consists of three parallel random forest regressors, each of them trained to calculate one of the stellar parameters considered. 
 This way, SPEEM only requires one run to receive the inputs (the 12 J-PLUS magnitudes) and to return \teff, \logg\, and \metal\  as outputs. 
  
  \subsection{Cleaning the sample from Extragalactic Sources and White Dwarfs}\label{morph}
  
 The estimation of stellar parameters requires a training sample free from contaminants  such as quasi-stellar objects (QSO) or white dwarfs (WD). These objects can increment the rate of false-positive candidates when it comes to searching for new VMP candidates. These unresolved contaminants might mimic VMP stars from photometric estimation, as shown in previous works using data from SEGUE \citep{dawson2012baryon} or from the Baryon Oscillation Spectroscopic Survey \citep[BOSS;][]{bolton2012spectral}. 

 For this purpose, in this work we built a data classifier   using the \texttt{SPECTYPE\_CLASS} flag provided by SEGUE as the target. We used the J-PLUS$\times$SEGUE dataset as the training sample to make a preliminary spectral type classification of the Gold sample, allowing us to identify and remove objects that may resemble VMP candidates in the J-PLUS color space.

 The separation between QSO and stars came from a machine learning model trained with photometric features with higher relative importance, selected by RF in a two-step process from a list of 12 J-PLUS magnitudes and 66 J-PLUS colors. 
  
 Although the WISE magnitudes lie in the infrared regime while J-PLUS magnitudes  correspond to the optical, the QSO-star separation of our sample was inspired by the  W1$-$W2$\times$W2$-$W3 diagram, presented by \citet{wright2010wide} and \citet{scaringi2013spectroscopic}, as a tool to separate QSOs  (with W1$-$W2$>$0.5) and stars (with W1$-$W2$<$0.5).
In the J-PLUS photometric system, we built a diagram based on the two essential and not correlated colors, (\textit{u}$-$J0410)$\times$(J0378$-$\textit{g}), from the RF feature importance list to separate QSOs from stars. 
An example of this Star-QSO separation applied to the sample J-PLUS$\times$SEGUE is shown on the left panel of Figure~\ref{fig:star_QSO}, suggesting there is a threshold at u$-$J0410$=0.5$ as a preliminary boundary to separate QSOs (represented as pink filled circles) from stars, color-coded according to the spectral type assigned by SEGUE in the insert. The right panel of Figure~\ref{fig:star_QSO} shows the distribuion of classes used in the training data ingested into the model.
  
\begin{figure}[]
    \centering
    \includegraphics[scale=0.7]{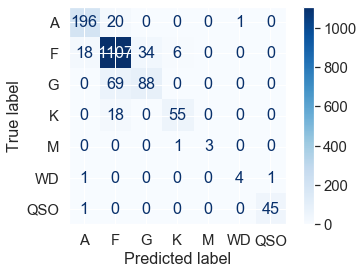}
    \caption{Confusion Matrix of the morphological and spectral type classification  applied to a test sample of 1\,668 objects from J-PLUS$\times$SEGUE not used in the training process. The blue bar indicates the number of objects of each class.}
    \label{spectral_class}
    \vspace*{-0.3cm}
  \end{figure}

Then the J-PLUS$\times$SEGUE sample was randomly split into two subsamples with a 0.75/0.25 ratio, giving a total of 5\,004 objects for training and 1\,668 objects for testing the predictions. Figure~\ref{spectral_class} presents the confusion matrix  of the SPEEM pipeline for the classification process, color-coded by the number of objects in each class. The  accuracy of the classifier, or the overall rate of correct classification, is 0.91. As shown in the confusion matrix presented in Figure~\ref{spectral_class}, the identification rate of QSOs and M stars was 98\% and 75\%, respectively. Spectral type stars A, F and K (196 out of 217, 1\,107 out of 1\,165  and 55 out of 73, respectively) presented a satisfactory recovery rate. On the  other hand, the lower metrics corresponding to G-stars (88 out of 157 stars correctly classified) is probably  due to the misclassification of  F and K type stars. Finally, the identification rate of WDs was around 67\%.

 At this stage of our analysis, we are interested in excluding the possible QSOs and WDs from our sample, resulting in an almost pure sample, ideal for training SPEEM to estimate stellar parameters.
  
  \section{Estimation of Parameters}
  \subsection{Training for Machine Learning Regressions}
  
\begin{figure*}[]
  \centering
  \begin{subfigure}[]{0.48\textwidth}
  \centering
  \includegraphics[width=\textwidth]{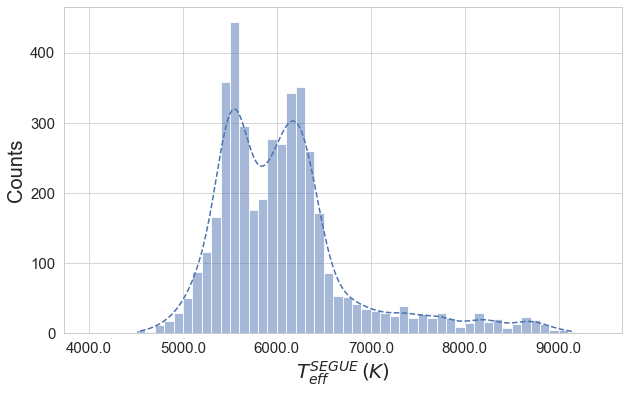}
  \end{subfigure}
  \hfill
  \begin{subfigure}[]{0.45\textwidth}
  \centering
  \includegraphics[width=\textwidth]{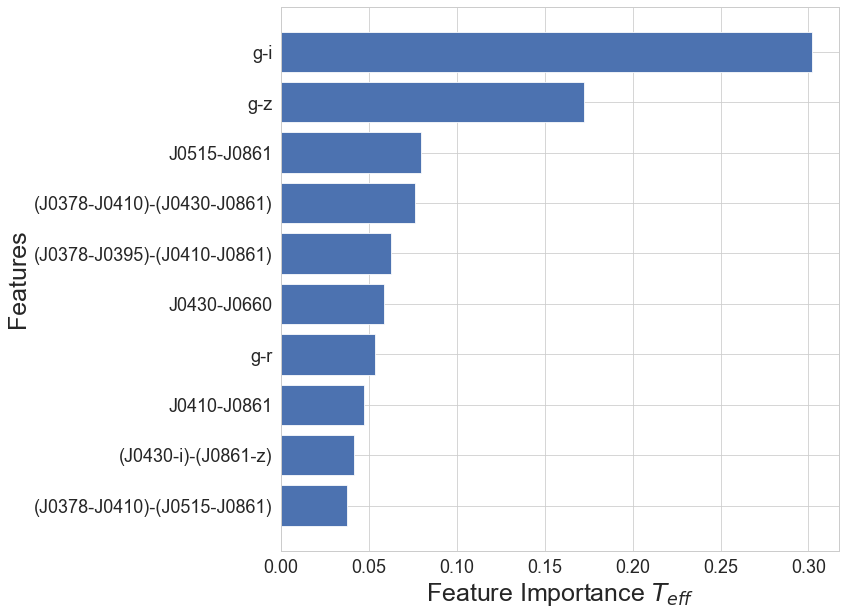}
  \end{subfigure}
  \caption{{\it Left:} Distribution of effective temperatures  for the training sample composed of 4,299 stars from J-PLUS$\times$SEGUE. {\it Right}: Relative importance of the features for the estimation of effective temperature. The most important features as \teff\ indicators are the colors \textit{g}-\textit{i}, \textit{g}-\textit{z}, and the color J05150$-$J0861 based on the narrow-band filters that contain the \ion{Mg}{i} Triplet and the \ion{Ca}{ii} Triplet.}
  \label{fig:teff_features}
  \end{figure*}

\begin{figure*}[htpb!]
  \centering
  \begin{subfigure}[]{0.48\textwidth}
  \centering
  \includegraphics[width=\textwidth]{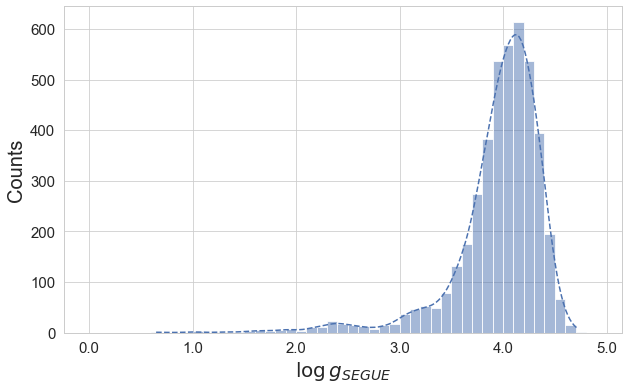}
  \end{subfigure}
  \hfill
  \begin{subfigure}[]{0.45\textwidth}
  \includegraphics[width=\textwidth]{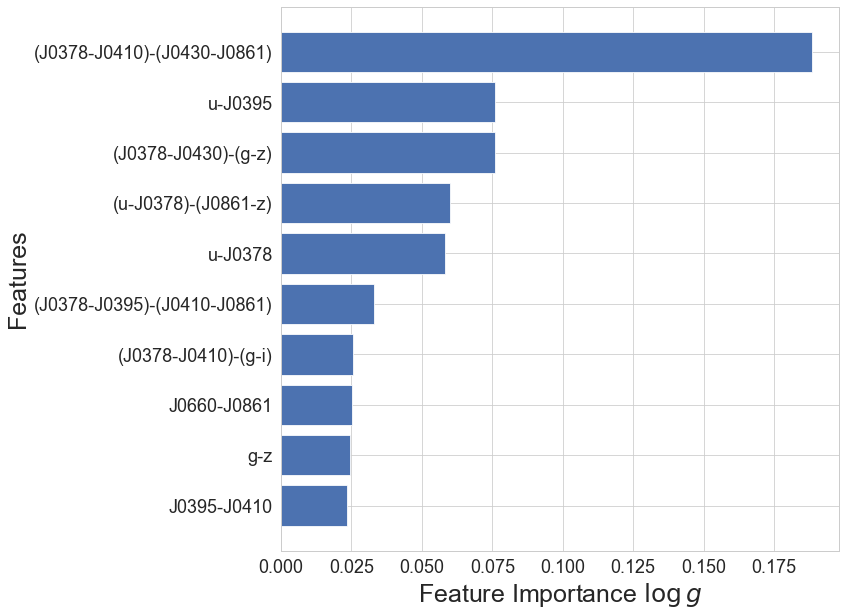}
  \end{subfigure}
  \caption{{\it Left:} Distribution of surface  gravities for the training sample composed of 4,299 stars from J-PLUS$\times$SEGUE. 
  {\it Right}: Relative importance of the features for the estimation of surface gravity. The most important feature as \logg\ indicator is the combination (J0378$-$J0410)$-$(J0430$-$J0861)}
  \label{fig:logg_features}
  \end{figure*}

\begin{figure*}[htpb!]
    \centering
    \begin{subfigure}[]{0.48\textwidth}
    \centering
    \includegraphics[width=\textwidth]{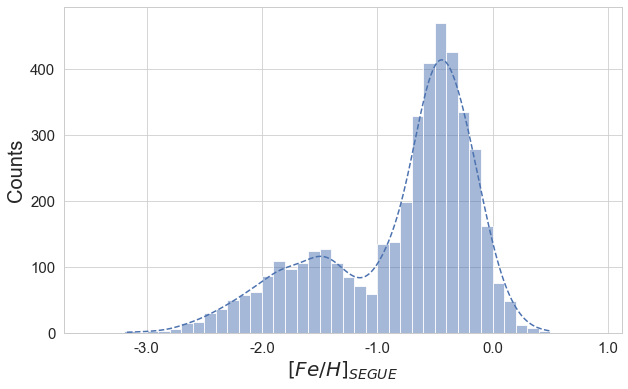}
    \end{subfigure}
    \hfill
    \begin{subfigure}[]{0.45\textwidth} \includegraphics[width=\textwidth]{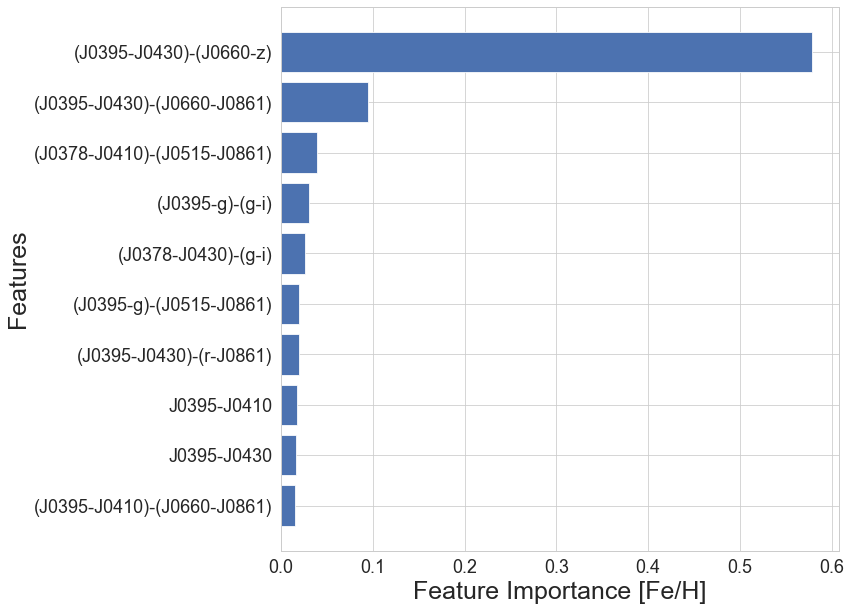}
    \end{subfigure}
    \caption{{\it Left:} Distribution of metallicity for the training sample composed of 4,299 stars from J-PLUS$\times$SEGUE. {\it Right}: Relative importance of the features for the estimation of metallicity. The most important feature as \metal\ indicator is the combination (J0395$-$J0430)$-$(J0660$-${\it z}) based on narrow-band filters containing the spectral features \ion{Ca}{ii} H \& K, G-band, and H$\alpha$ and the broad-band filter {\it z}.}
     \label{fig:feh_features}
  \end{figure*}

  The estimation of \teff, \logg, and \metal\ was based on a training sample with 4\,299 stars with spectral types F and G, randomly selected from J-PLUS$\times$SEGUE (after removing QSOs, WDs and bad flags spectra) described in Section~\ref{morph}. The training sample presents a distribution of effective temperatures between 4\,800~K and 9\,000~K; surface gravities between 1.0 and 4.5, and metallicities between $-$3.0 and $+$0.5 estimated by SEGUE, as shown in the left panels of the Figures~\ref{fig:teff_features} to \ref{fig:feh_features}. The distribution of \teff\ (Fig. \ref{fig:teff_features}) is consistent with the temperatures corresponding to the selected spectral types F and G, with possible contamination of some misclassified A stars, that extends the \teff\ distribution up to 9\,000~K. Figure~\ref{fig:logg_features} indicates that main-sequence stars dominate the training sample but this also contains a few evolved stars. The metallicity distribution in Figure~\ref{fig:feh_features} suggests the training sample contains stars of the thick disk, corresponding to the peak at \metal\ $\sim -0.6$, as well as inner halo stars, corresponding to \metal\ $\sim -1.6$ \citep{carollo2010}.

In this work, we are mainly interested in deriving the atmospheric parameters and searching for new candidates for VMP stars, so the photometric estimation of \metal\ is crucial. However, the determination of stellar metallicity based on photometric measurements is especially challenging at the resolution of J-PLUS data: for a given temperature, the contrast between the metal absorption features and the continuum decreases significantly at lower metallicity values. Thus, it is crucial to define a training sample covering a wide range of \metal\ values. For this reason,  the J-PLUS$\times$SEGUE dataset, which contains stars of \metal\ between $-3.17$ and $0.5$ in a bimodal distribution, as shown in Figure~\ref{fig:feh_features}, is the most adequate for training the model.

  The statistical correlation analysis of the complete set of colors and all possible combinations of J-PLUS filters with the stellar parameters \teff, \logg\ and \metal\,  allows us to select the best set of features, i.e., those features with  the highest Pearson's correlation coefficient, to feed RF in order to obtain the best accurate predictions.
  RF provides the relative importance assigned to input features for each estimated parameter (as shown in the right panels of Figures~\ref{fig:teff_features} to ~\ref{fig:feh_features}) as feedback to test the model performance and the precision of the estimations. The feature importance
  score corresponds to the average of each decision tree within the trained model \citep{hastie2009elements}. For example, in the case of \teff, the essential features  include the colors {\it g}$-${\it i} and {\it g}$-${\it z}  and the color   J0515$-$J0861, based on the narrow-band filters centered on the \ion{Mg}{i} triplet and the \ion{Ca}{ii} triplet, respectively. For \logg, the best indicator is the color (J0378$-$J0410)$-$(J0430$-$J0861),
  that contains the spectral features H$\delta$, G-band, and the \ion{Ca}{ii} triplet. Finally, for \metal, the index (J0395$-$J0430)$-$(J0660$-${\it z}), based on filters containing the spectral features \ion{Ca}{ii} H \& K, G-band, and H$\alpha$, excels as a good metallicity indicator. 
  
  \begin{figure}[htpb!]
    \centering
   \includegraphics[scale=0.27]{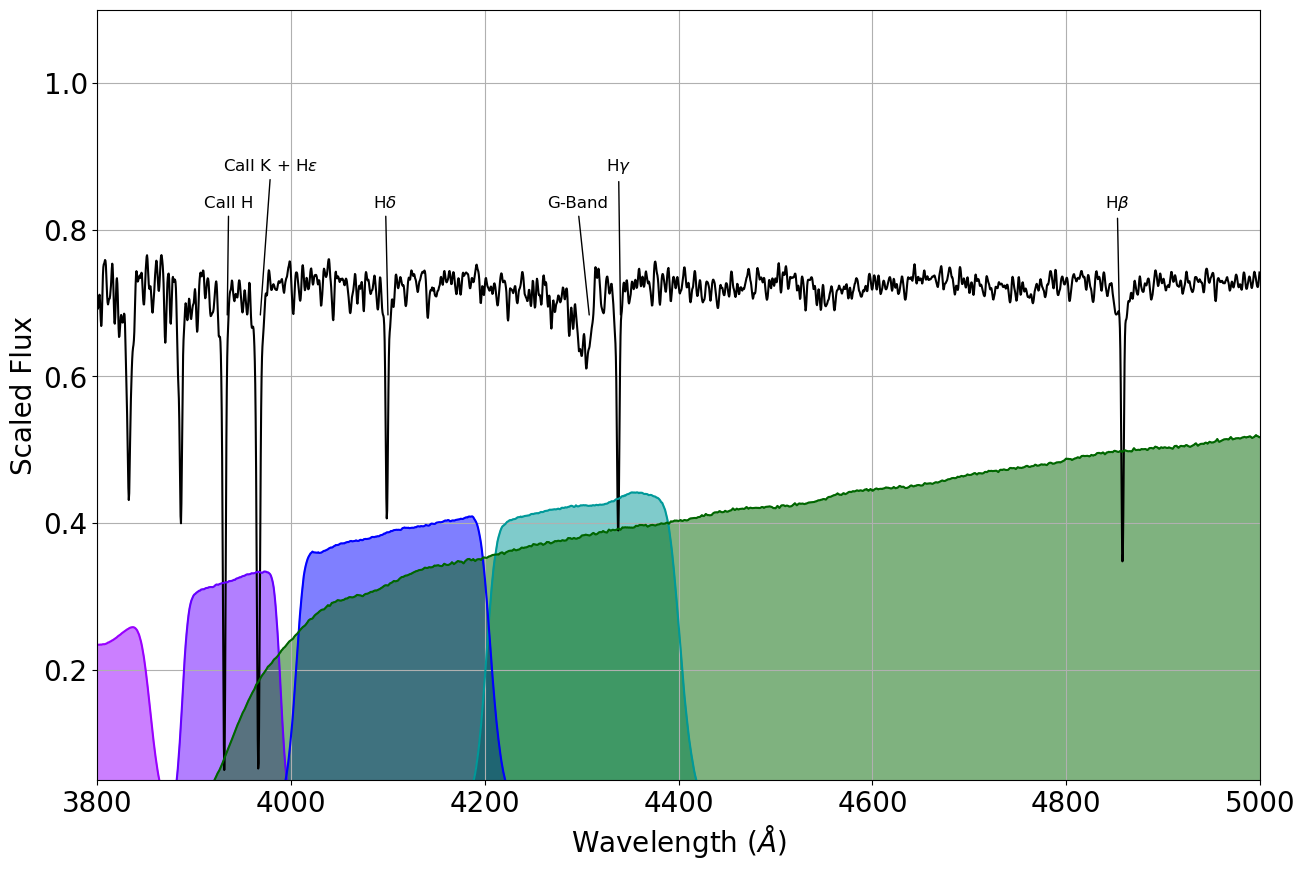}
    \caption{Some J-PLUS filters overplotted to the spectrum of the G-type star J-PLUS 75091-15989 with \metal$=-3.09$. The filters J0395 and J0410 contain the \ion{Ca}{ii} H \& K and the Hydrogen line H$\delta$, respectively. The H$\gamma$ line and the G-band  are evident in the filter J0430.}
    \label{specfilter}
    \vspace{-0.3cm}
  \end{figure}
  
  Figure~\ref{specfilter} shows the J-PLUS filters J0395, J0410, J0430, and  \textit{g} overplotted to the observed spectrum of a G star (we refer to Section~\ref{candidates} for more details on the observations and reduction of the spectroscopic data),   with 
  the identification of some essential spectral features such as \ion{Ca}{ii} H \& K lines, the Balmer lines H$\beta$, H$\gamma$, H$\delta$, and the G-band.
  These spectral features, however, may be sensitive to more than one atmospheric parameter simultaneously, introducing a degeneracy in determining the parameters. For example, Figure 4 of \citet{whitten2019j} exhibits the sensitivity of the \ion{Ca}{ii} H \& K lines with \teff\ and \metal. On the other hand, H lines are mainly sensitive to effective temperature for stars with spectral types later than F. Thus, the analysis based on a combination of filters containing spectral features with different sensitivity to the atmosphere parameters may resolve this degeneracy.
 
 \begin{figure*}[hbt!]
     \centering
     \begin{subfigure}[b]{0.3\textwidth}
     \centering
     \includegraphics[width=\textwidth]{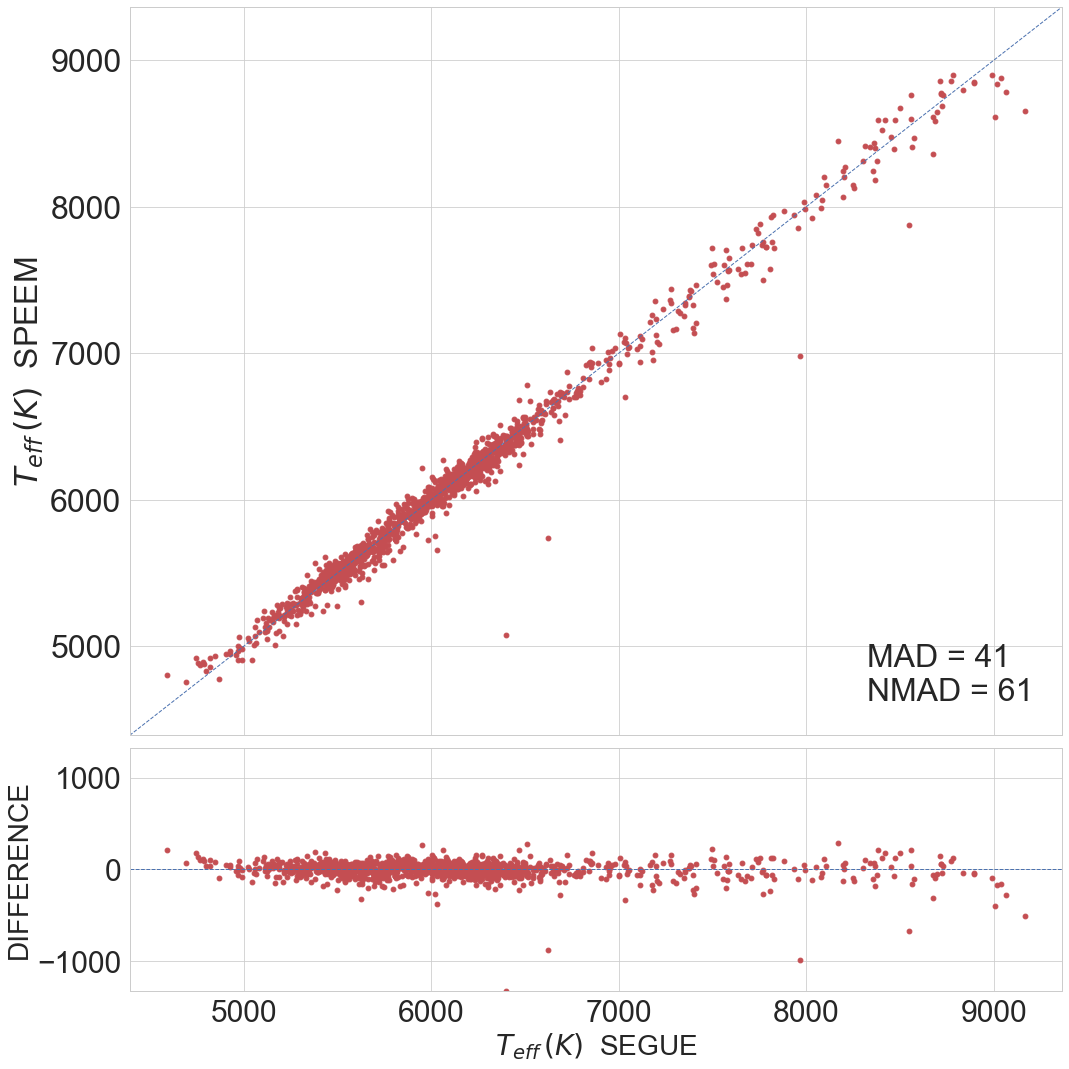}
     \end{subfigure}
     \hfill
     \begin{subfigure}[b]{0.3\textwidth}
     \centering
     \includegraphics[width=\textwidth]{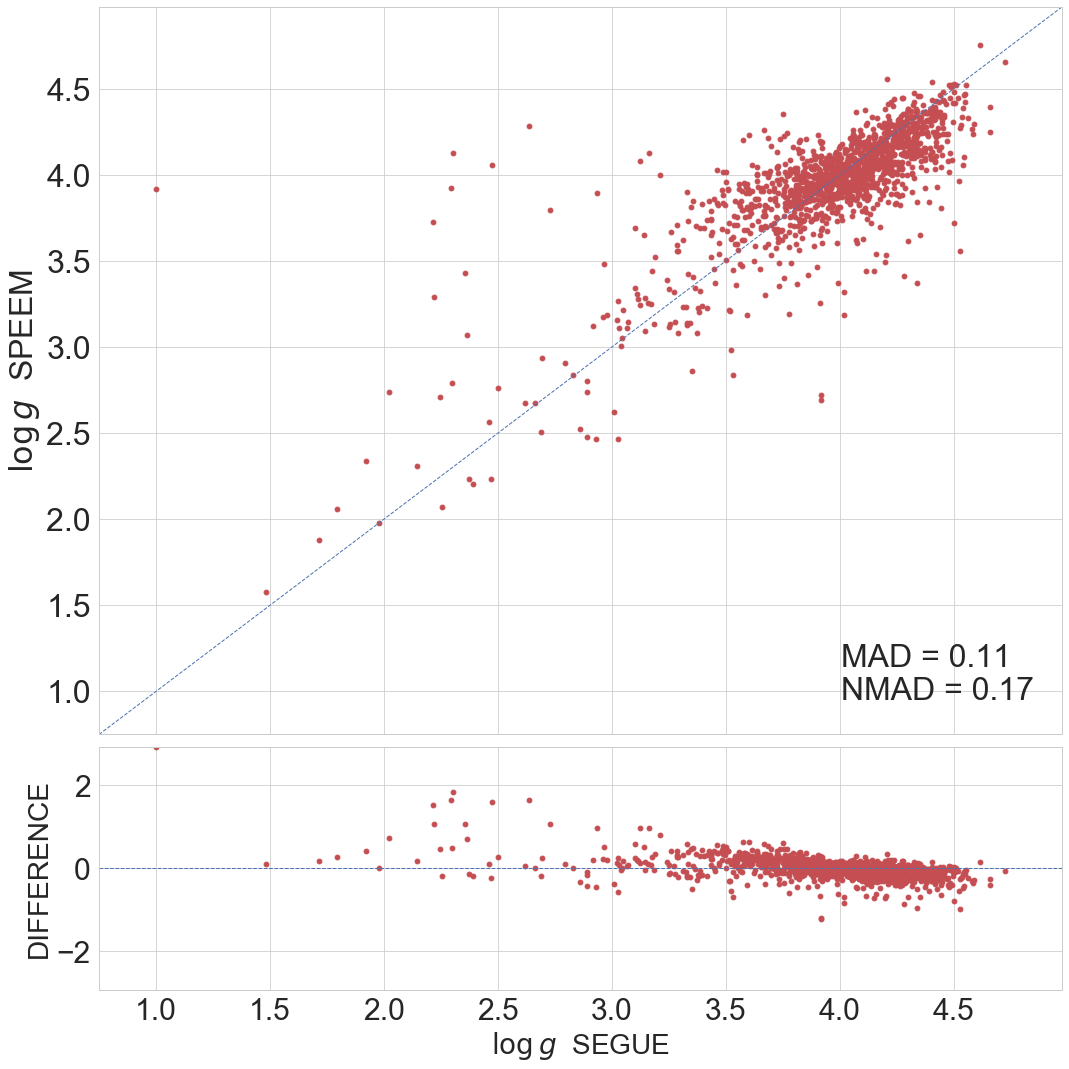}
     \end{subfigure}
     \hfill
     \begin{subfigure}[b]{0.3\textwidth}
     \centering
     \includegraphics[width=\textwidth]{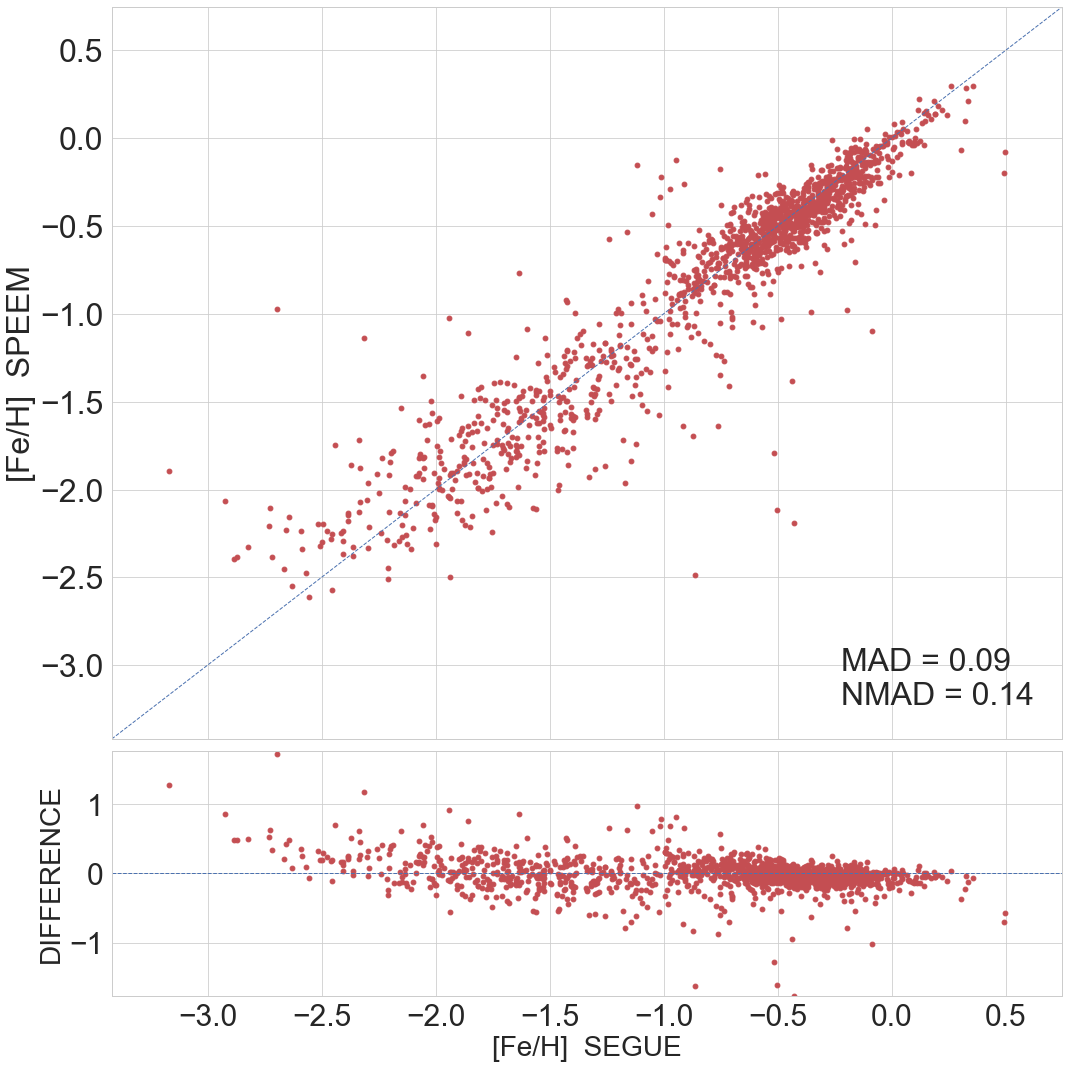}
     \end{subfigure}
    \hfill
    \begin{subfigure}[b]{0.3\textwidth}
     \centering
     \includegraphics[width=\textwidth]{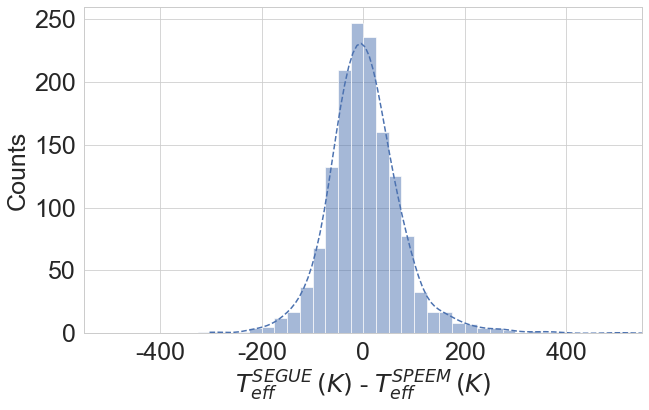}
     \end{subfigure}
     \hfill
     \begin{subfigure}[b]{0.3\textwidth}
     \centering
     \includegraphics[width=\textwidth]{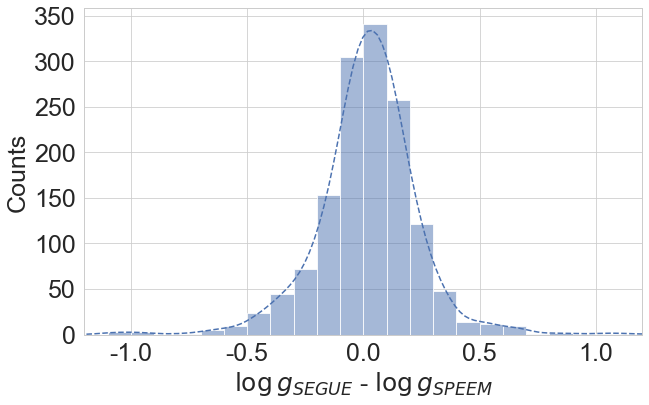}
     \end{subfigure}
     \hfill
     \begin{subfigure}[b]{0.3\textwidth}
     \centering
     \includegraphics[width=\textwidth]{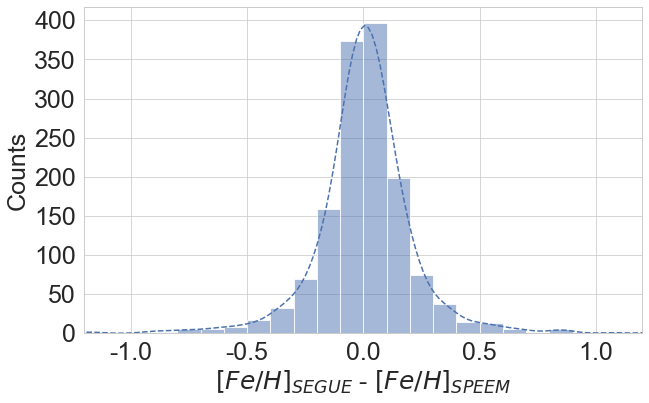}
     \end{subfigure}
     \caption{Comparison between  parameters  \teff\ ({\it upper left panel}), \logg\ ({\it upper middle panel}) and \metal\ ({\it upper right panel}) obtined  with SSPP and  the SPEEM pipelines for $0.25\%$ of the sample J-PLUS$\times$SEGUE acting as a test sample. The median absolute deviation (MAD) and the corresponding normalized median absolute deviation (NMAD) are shown in each panel. The lower panels show the distributions of the differences between the respective parameters obtained with  SSPP and SPEEM.     
     }
      \label{fig:comp_segue}
  \end{figure*} 

   \begin{figure*}[]
   \centering
   \begin{subfigure}[]{0.3\textwidth}
   \centering
   \includegraphics[width=\textwidth]{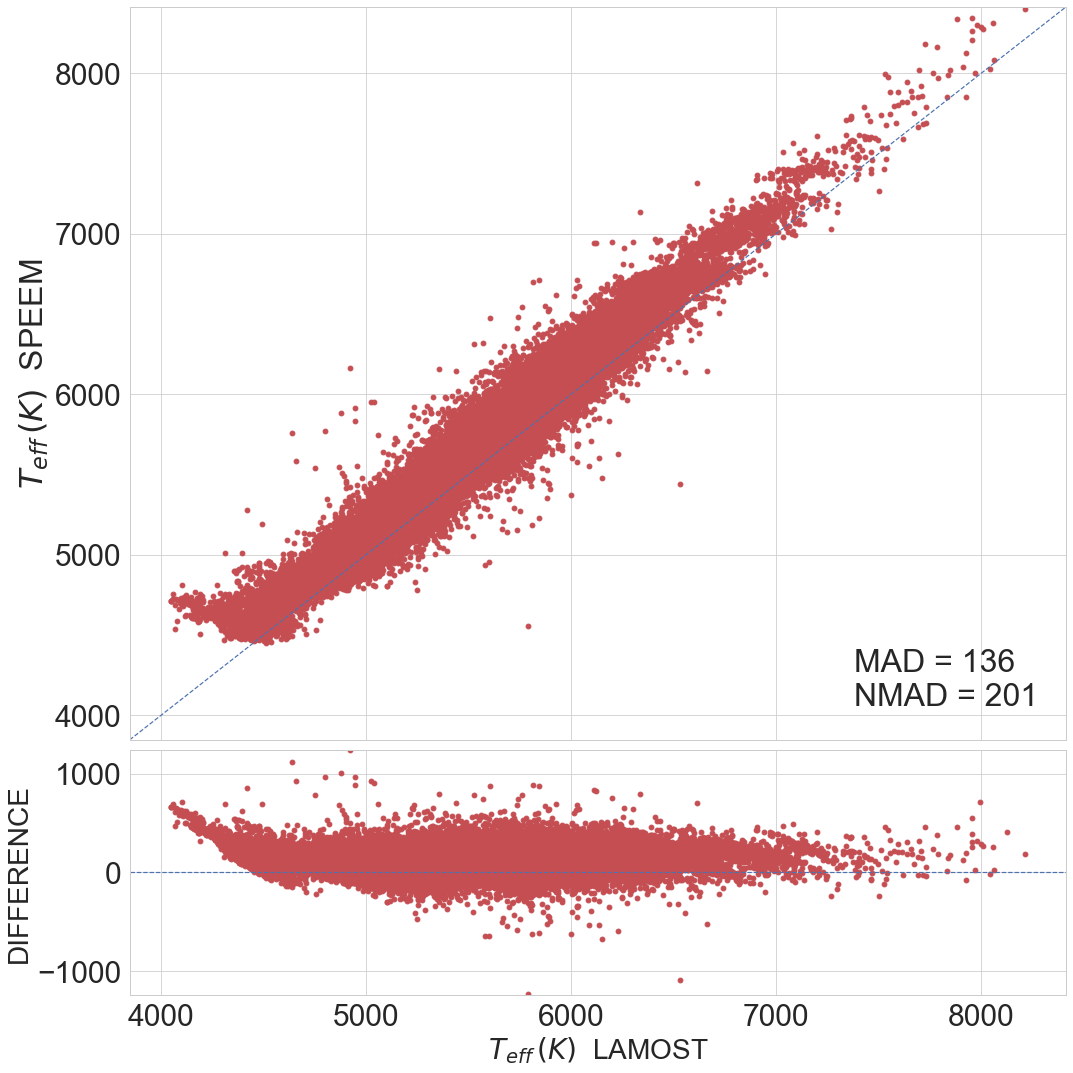}
   \end{subfigure}
   \hfill
   \begin{subfigure}[]{0.3\textwidth}
   \centering
   \includegraphics[width=\textwidth]{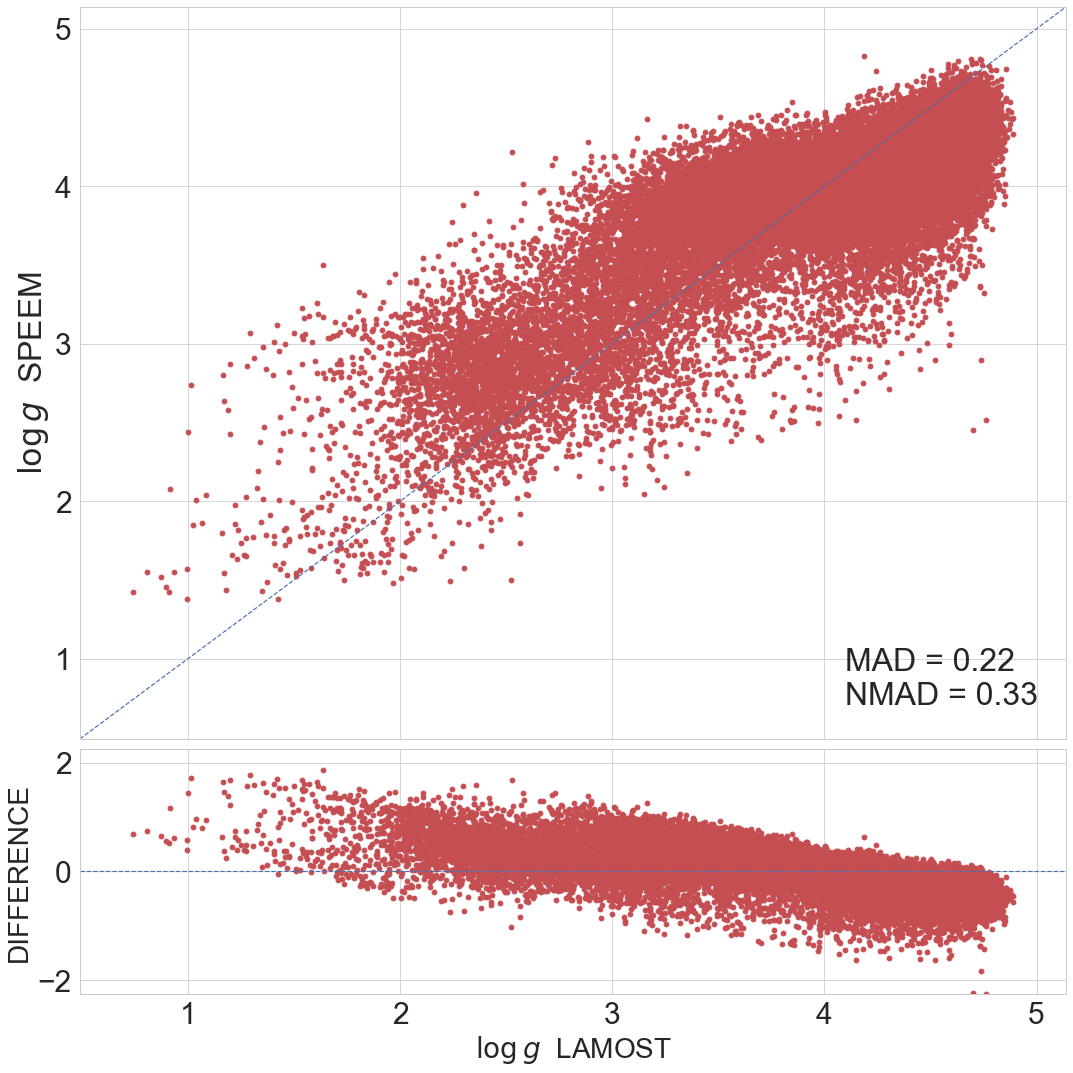}
   \end{subfigure}
   \hfill
   \begin{subfigure}[]{0.3\textwidth}
   \centering
   \includegraphics[width=\textwidth]{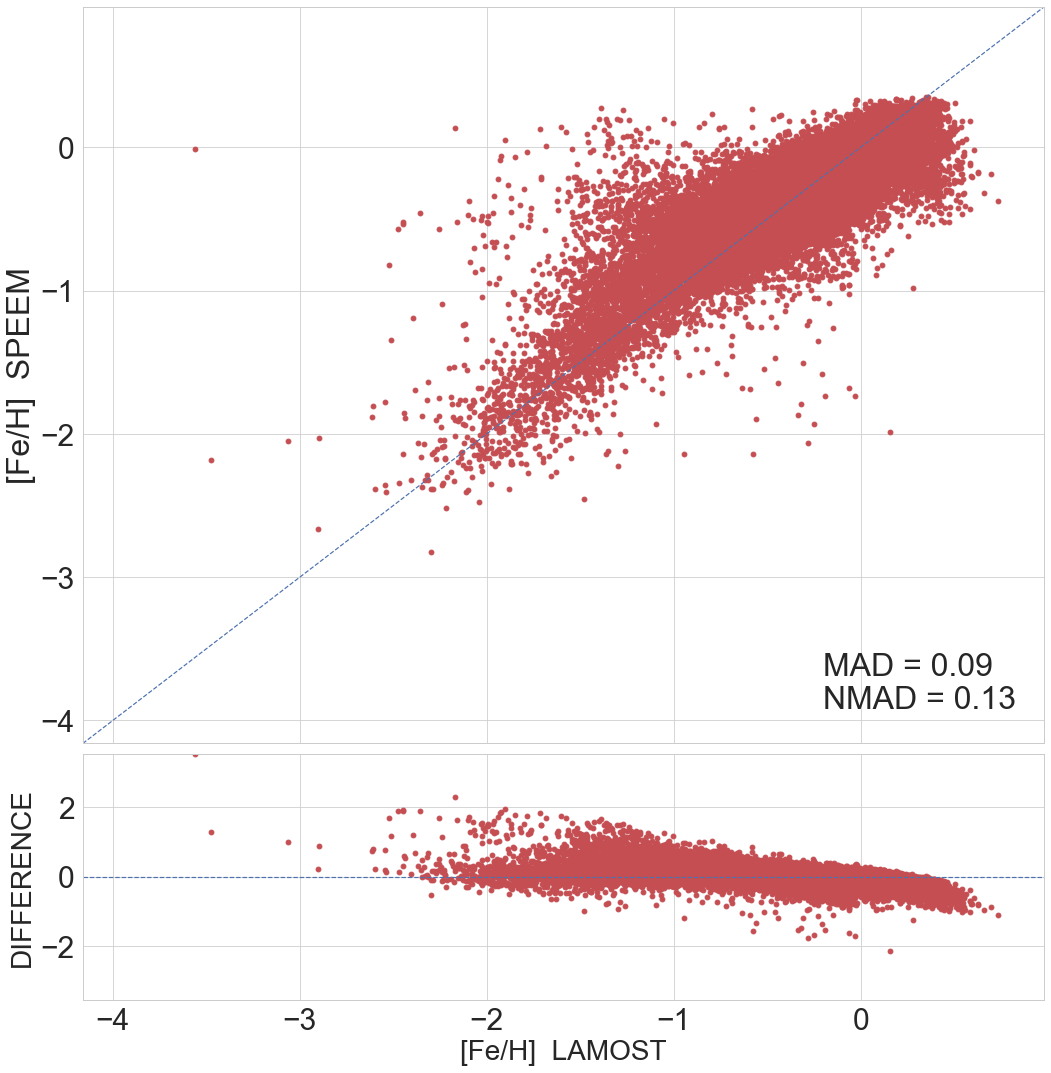}
   \end{subfigure}
   \hfill
    \begin{subfigure}[]{0.3\textwidth}
     \centering
     \includegraphics[width=\textwidth]{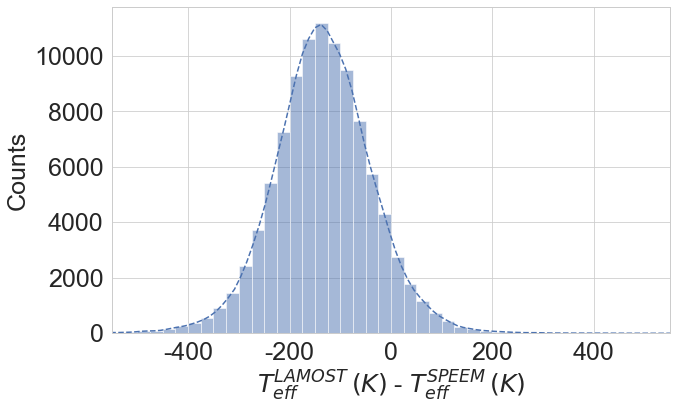}
     \end{subfigure}
     \hfill
     \begin{subfigure}[]{0.3\textwidth}
     \centering
     \includegraphics[width=\textwidth]{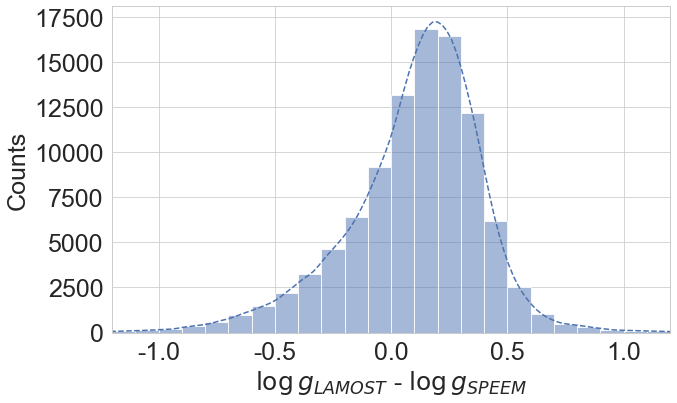}
     \end{subfigure}
     \hfill
     \begin{subfigure}[]{0.3\textwidth}
     \centering
     \includegraphics[width=\textwidth]{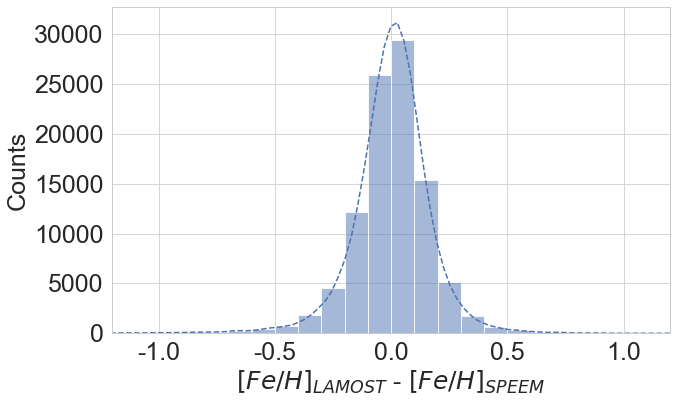}
     \end{subfigure}
    \caption{Comparison between  parameters  \teff\ ({\it upper left panel}), \logg\ ({\it upper middle panel}) and \metal\ ({\it upper right panel}) obtained  with n-SSPP and the SPEEM pipeline for stars in of the sample J-PLUS$\times$LAMOST acting as a validation sample. The mean absolute error (MAD) and the corresponding  standard deviation are shown in each panel. The lower panels show the distributions obtained of the differences between the respective parameters obtained with  n-SSPP and SPEEM.}
    \label{fig:comp_lamost}
  \end{figure*} 

\subsection{Validation of Stellar Parameters}

Once the different SPEEM models completed the learning process, we 
tested the pipeline on  a subsample of 1\,668 stars  (corresponding to 25\% of the J-PLUS$\times$SEGUE sample not used in the training process) in order to compare the values estimated by SEGUE (medium-resolution spectroscopy) with the ones predicted by SPEEM as shown in Figure~\ref{fig:comp_segue}. The accuracy of the predictions correspond to the median absolute deviation (\textrm{MAD}), and the normalized absolute deviation (\textrm{NMAD}). 

The \teff\ values estimated by SPEEM are in good agreement with the ones estimated by the pipeline SSPP of SEGUE (upper left panel of Figure~\ref{fig:comp_segue}), obtaining an average error of $41\pm61$~K. Most of the stars in the test sample have \teff\ $<$ 7\,000~K,  since it is highly dominated by F-type stars (Section~\ref{morph}, right panel of Figure~\ref{fig:star_QSO}). The histogram in the lower left panel of Figure~\ref{fig:comp_segue}) represents the distribution of the differences between the \teff\ obtained with SSPP and SPEEM, and it shows no systematic difference between the two temperature scales.

The middle panel of Figure~\ref{fig:comp_segue} shows the correlation between SSPP values and SPEEM predictions for \logg\, with MAD = 0.11$\pm$0.17~dex. The \logg\ values estimated by SPEEM agree with those derived by SSPP within $0.05$~dex for main-sequence stars. On the other hand, the mean difference is $0.26$~dex for more evolved stars, with \logg\ $\leq3.5$. Even the use of different colors as surface gravity indicators
(Figure~\ref{fig:logg_features})
does not provide overall precision high enough to make more accurate estimations of \logg. The study of the photometric determination of surface gravity anchored on asteroseismology is ongoing and will be the subject of a future paper.

Finally, the upper right panel of Figure~\ref{fig:comp_segue} presents a reasonably good correlation between \metal\ values derived by SSPP and SPEEM for \metal$> -1.0$. However, there is a more significant dispersion in the lower metallicity regime. The MAD and NMAD values are $0.09\pm0.14$~dex, and the differences show a slight slope for \metal$<-2.0$, suggesting that SPEEM may overestimate the \metal\ value in the lower range of the metallicity distribution. This trend is probably due to the metallicity distribution and the lack of data required for a proper learning process in this metallicity range. 
 
An additional test of the SPEEM capabilities has been performed with the J-PLUS$\times$LAMOST database and using the same training sample of 4\,299 stars described in the previous section. The upper panels of Figure~\ref{fig:comp_lamost} show the comparisons between stellar parameters estimated by SPEEM and those obtained with n-SSPP applied to LAMOST spectra.

The typical scatter between the \teff\ values from the two datasets is lower than $\sim$150 K. The dispersion is higher than in comparison with SEGUE shown in Figure~\ref{fig:comp_segue}, although the range of \teff\ is extended towards lower temperatures, reaching 4\,500~K. The estimation of surface gravity shows a larger dispersion with an average difference of 0.22$\pm$0.33~dex while the comparison for metallicity has MAD = 0.09$\pm$0.13. The distributions of the differences between the databases are shown in the respective lower panels. The difference \metal$_{\textrm{SPEEM}}-$\metal$_\textrm{{LAMOST}}$ shows a trend with \teff$_{\textrm{,SPEEM}}-$\teff$_{\textrm{{,LAMOST}}}$ so that a change of $100$~K in \teff\ produces a variation of $\sim0.2$~dex in \metal. 

An extra test using a larger sample resulting from SEGUE and LAMOST merged databases to train the entire spectroscopic parameter space shows that such combination adds many stars with \metal $>-2.5$ but has no significant contribution in the lower end of the metallicity distribution. As a result, a combined training sample reinforces the unbalance in the metallicity distribution and introduces more considerable uncertainty in the metallicity estimation. For this reason, we decide to keep the J-PLUS$\times$SEGUE as the training sample.
   
  \begin{figure*}[]
  \begin{subfigure}[]{0.48\textwidth}
  \centering
  \includegraphics[width=\textwidth]{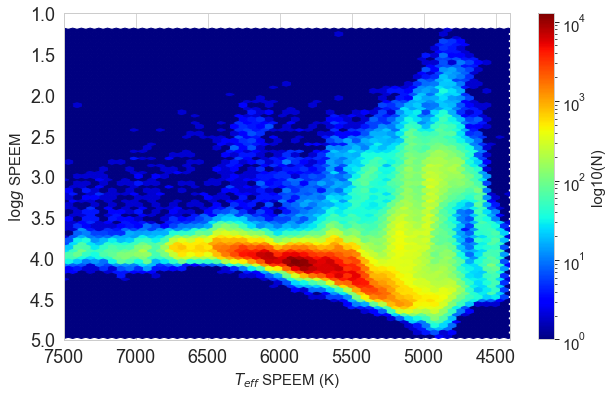}
  \end{subfigure}
  \hfill
  \begin{subfigure}[]{0.48\textwidth}
  \includegraphics[width=\textwidth]{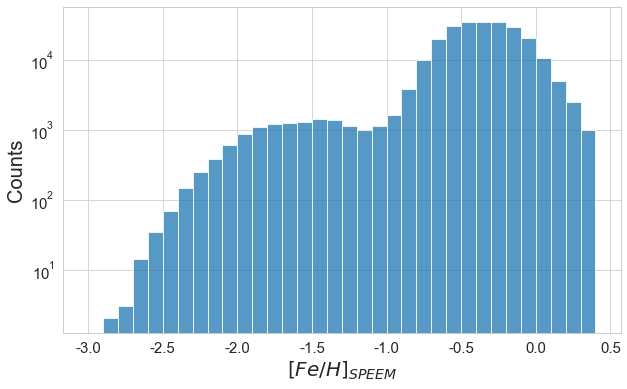}
  \end{subfigure}
  \caption{{\it Left:} The distribution of stellar parameters \teff\ and \logg\ obtained with SPEEM for the Gold sample. The points are color-coded according to the number of stars, indicating the sample is mostly composed by main sequence stars with spectral types F and G. 
  {\it Right}: The distribution of \metal\ obtained with SPEEM for the Gold sample. The distribution has two peaks related to the stellar populations of the thick disk at \metal$\sim-$0.4 and the halo at \metal$\sim-$1.6}. 
  \label{hr}
  \vspace*{-0.3cm}
  \end{figure*}

  \section{Searching for VMP candidates}\label{candidates}
  
  After completing the validation process, we applied SPEEM to the entire Gold Sample and obtained effective temperatures, surface gravities, and metallicities for the 746\,531 stars. The left panel of Figure~\ref{hr} shows the Hertzsprung-Russell Diagram for the Gold Sample, with the stellar parameters \teff\ and \logg\ obtained with SPEEM. The points are color-coded according to the number of stars in the parameters space and the observed distribution suggests that the main sequence stars dominate the Gold sample with spectral types F and G.
  
  The right panel of Figure~\ref{hr} shows the metallicity distribution obtained for the Gold Sample. The histogram has two peaks related to the stellar populations of the thick disk at \metal$\sim -0.4$ and the halo at \metal$\sim -1.6$. This distribution follows the general metallicity distribution of the halo widely discussed in the literature \citep[e.g.,][]{carollo2007} associated with different components in the Galactic Halo. 

  A subsample of 575\,600 objects 
  was used for our search for new VMP candidates. In order to define a list for a spectroscopic follow-up, we imposed the following selection criteria: \teff $<$ 5\,500 K (corresponding to spectral types later than $\sim$G) and  \metal$<-2.5$. The limit magnitude has been   constrained  even further (from  \textit{g} $<$ 18  to \textit{g} $<$ 17) for observations with 4-m class telescopes. The application of these selection criteria produced a list of 177 low metallicity candidates.
  
  From this list, we selected 11 stars with coordinates optimal for observations with the 4.2-m William Herschel Telescope at the Roque de Los Muchachos Observatory, La Palma.  The observations   with the WHT coupled to the spectrograph ISIS with a 1$\arcsec$ slit provide spectra (in the blue arm) in the range 3900--5100\,\AA\, and nominal resolution $\lambda/\Delta\lambda \sim$ 2\,000 at 4000\,\AA.  
  In Table~\ref{table:2} we list the observational data for the sample of candidates: the equatorial coordinates, observation dates, exposure times, {\it g} magnitudes, and the measured signal-to-noise ratios at $\sim 4\,500$ \AA.

  The data reduction followed the standard procedure of bias subtraction, flat-field correction, extraction of the one-dimensional spectra, and wavelength calibration using the IRAF\footnote{IRAF was distributed by the National Optical Astronomy Observatories, which are operated by the Association of Universities for Research in Astronomy, Inc., under cooperative agreement with the National Science Foundation.} software.
  
  The stellar atmospheric parameters were determined using the n-SSPP \citep{Beers_2014,beers2017}, providing the \teff, \logg, and \metal\ listed in Table~\ref{table:3}. Also listed are the estimates obtained with SPEEM for these parameters and effective temperatures from the Gaia DR2 Catalogue. The n-SSPP uses as input photometric information and the observed spectrum for each star. The parameters are determined based on photometric calibrations, line indices, and matching with a database of synthetic spectra. Further details on the procuderes can be found on \citet{lee2008seguea,lee2008segueb}.
    
\begin{figure*}
   \centering
   \includegraphics[scale=0.68]{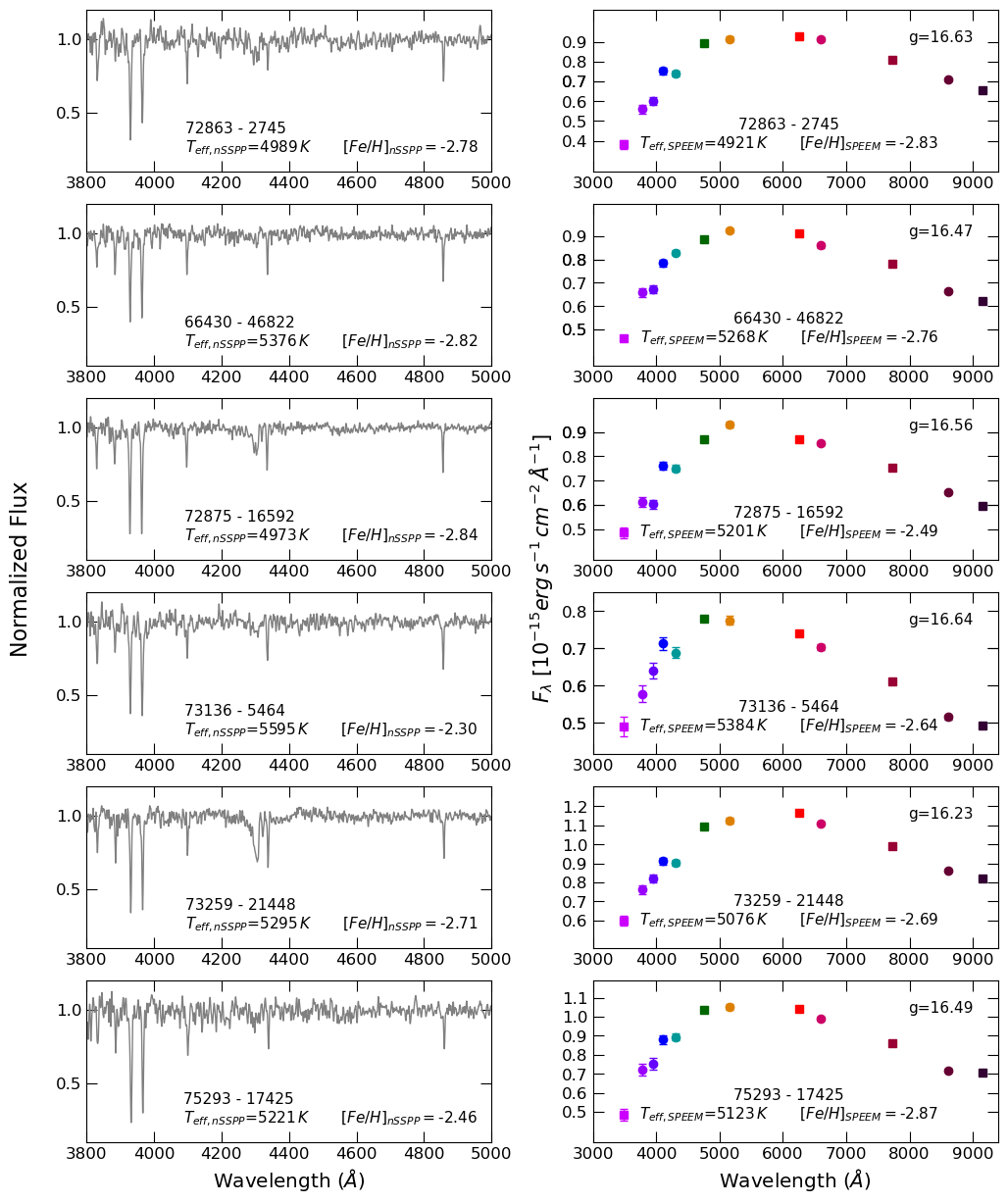}
   \caption{{\it Left panels}: WHT/ISIS spectra of  very metal-poor candidates identified according to Table~\ref{table:2}.  
   The \ion{Ca}{ii} H \& K lines and the  Hydrogen lines H$\beta$, H$\gamma$, and H$\delta$, as well as the G-band,  are clearly seen. {\it Right panels}: J-PLUS photo-spectra of the very-metal poor candidates. The points are color coded according to the J-PLUS filters; squares represent the broad band filters and circles represent the narrow band filters. }
   \label{photospec1}
   \vspace*{-0.3cm}
   \end{figure*}

\begin{figure*}
   \centering
   \includegraphics[scale=0.68]{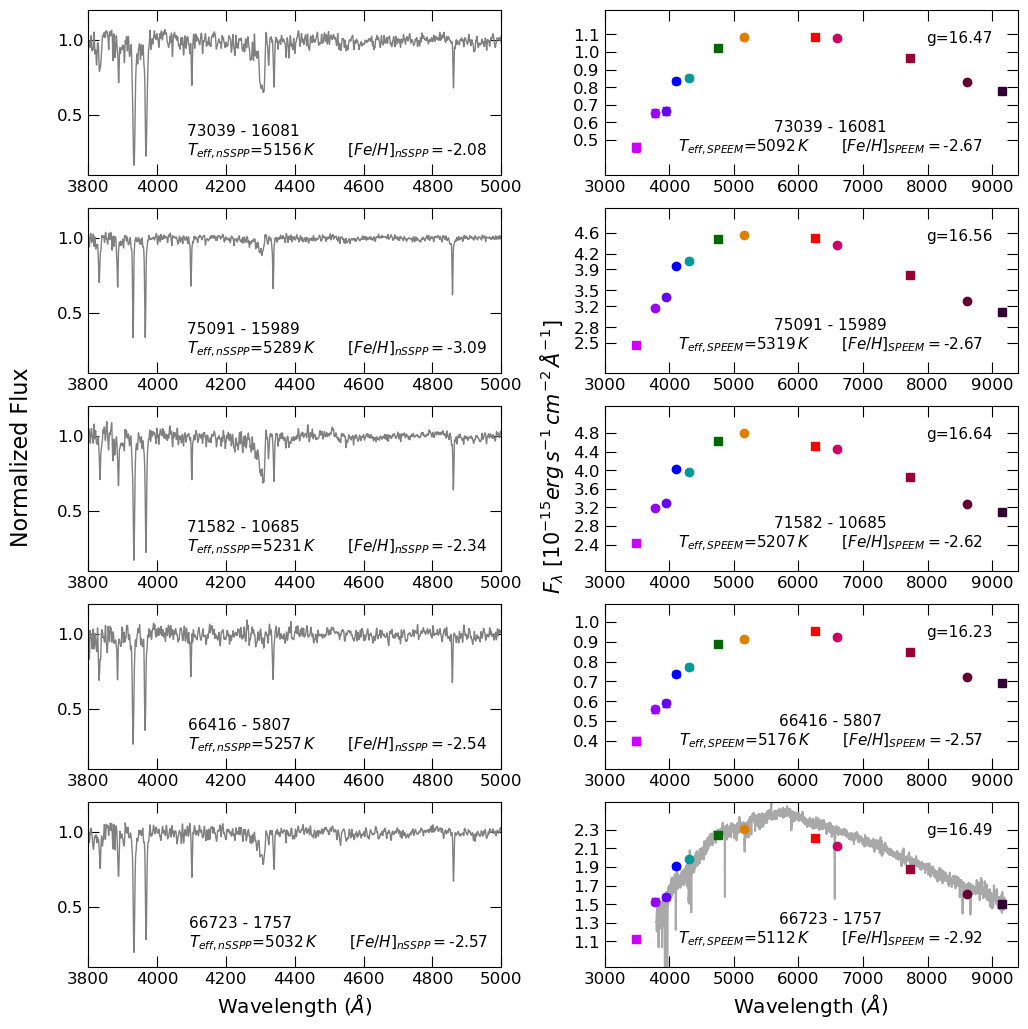}
   \caption{{\it Left panels}: WHT/ISIS spectra of  very  metal-poor candidates identified according to Table~\ref{table:2}.  
   The \ion{Ca}{ii} H \& K lines and the  Hydrogen lines H$\beta$, H$\gamma$, and H$\delta$, as well as the G-band,  are clearly seen. {\it Right panels}: J-PLUS photo-spectra of the very-metal poor candidates. The points are color coded according to the J-PLUS filters; squares represent the broad band filters and circles represent the narrow band filters. The gray line represents the SEGUE spectrum of the star 66723-1757 superposed to its photo-spectrum.}
   \label{photospec2}
   \vspace*{-0.3cm}
   \end{figure*}

  The left panels of Figures~\ref{photospec1} and ~\ref{photospec2} show the observed spectra obtained for the metal-poor candidates. The values of \teff\ and \metal\ obtained with the n-SSPP are indicated in each panel.  
  The right panels of Figures ~\ref{photospec1} and ~\ref{photospec2} show the J-PLUS photo-spectra for the sample stars. The Spectral Energy Distribution of the selected stars are consistent with the SED of a typical G-type star. The points are color coded according to the J-PLUS filters, as in \citet{JPLUS}. Squares represent the broad band filters and circles represent the narrow band filters. The medium-resolution spectrum obtained by SEGUE is available for only one star of our sample, J-PLUS ID 66723-1757, represented by the gray line in the lower right panel of Figure~\ref{photospec2}. Estimates of \teff\ and \metal\ obtained with SPEEM are indicated in each panel.
    
  \begin{table*}[]
  \centering
  \caption[]{\label{table:2}
  {Very metal-poor stars candidates: Spectroscopic data}}
  \begin{tabular}{lllllrl} \hline
  \hline
J-PLUS ID & RA (2000) & DEC (2000) & Date & {\it g} & t$_{\textrm {exp}}$ (s) &  S/N \\
    \hline
66416-5807  & 18:27:22.020 & +41:03:51.51 & 2019-09-28 & 16.36 & 1800.0 & 40 \\
66430-46822 & 18:37:21.190 & +42:04:11.34 & 2019-09-28 & 16.47 & 1800.0 & 55 \\
66723-1757  & 15:58:17.750 & +42:22:39.39 & 2019-09-28 & 15.66 & 1200.0 & 55 \\
71582-10685 & 00:13:06.280 & +02:30:46.71 & 2019-09-29 & 14.85 &  900.0 & 70 \\
72863-2745  & 01:34:37.850 & +06:54:11.02 & 2019-09-29 & 16.63 & 2200.0 & 40 \\
72875-16592 & 01:35:24.540 & +07:40:09.70 & 2019-09-28 & 16.56 & 1800.0 & 70 \\
73039-16081 & 22:11:14.900 & +10:34:53.30 & 2019-09-28 & 16.28 & 1800.0 & 60 \\
73136-5464  & 22:54:43.580 & +09:50:23.51 & 2019-09-29 & 16.64 & 1800.0 & 45 \\
73259-21448 & 22:48:39.880 & +12:11:39.52 & 2019-09-29 & 16.23 & 1800.0 & 60 \\
75091-15989 & 02:35:22.300 & +43:27:51.80 & 2019-09-28 & 14.66 &  600.0 & 90 \\
75293-17425 & 14:57:17.890 & +54:12:10.54 & 2019-09-28 & 16.49 & 1200.0 & 40 \\
\hline
   \end{tabular}
   \end{table*}

 \begin{table*}[]
 \centering
 \caption[]{\label{table:3}
 {Very metal-poor stars candidates: Stellar parameters obtained from SPEEM and from the  spectroscopic analysis with n-SSPP}}
 \begin{tabular}{c|c|ccc|ccc} \hline
 \hline
  J-PLUS ID &  \teff \ (K)  &  \teff \ (K) & \logg & \metal & \teff \ (K) & \logg & \metal \\  & \multicolumn{1}{c|}{Gaia} 
   & \multicolumn{3}{c|}{SPEEM} & \multicolumn{3}{c}{n-SSPP}  \\
  \hline
66416-5807 & 4\,994 & 5\,176 & 1.94 & $-2.57$ & 5\,257$\pm$55 & 2.82$\pm$0.55 & $-2.54\pm$0.12 \\
66430-46822 & 5\,038 & 5\,268 & 2.18 & $-2.76$ & 5\,376$\pm$87 & 3.45$\pm$0.46 & $-2.82\pm$0.13 \\
66723-1757 & 5\,058 & 5\,112 & 2.37 & $-2.92$ & 5\,032$\pm$69 & 2.59$\pm$0.21 & $-2.57\pm$0.14 \\
71582-10685 & 5\,105 & 5\,207 & 2.52 & $-2.62$ & 5\,231$\pm$50 & 1.72$\pm$0.10 & $-2.34\pm$0.10 \\
72863-2745 & 4\,936 & 4\,921 & 1.47 & $-2.83$ & 4\,989$\pm$44 & 1.12$\pm$0.21 & $-2.78\pm$0.18 \\
72875-16592 & 4\,860 & 5\,201 & 2.62 & $-2.49$ & 4\,973$\pm$103  & 2.29$\pm$0.55 & $-2.84\pm$0.15\\
73039-16081 & 4\,911 & 5\,092 & 1.81 & $-2.67$ & 5\,156$\pm$48 & 2.43$\pm$0.45 & $-2.08\pm$0.09 \\
73136-5464 & 5\,125 & 5\,384 & 2.71 & $-2.64$ & 5\,595$\pm$50 & 2.80$\pm$0.23 & $-2.30\pm$0.14 \\
73259-21448 & 4\,968 & 5\,076 & 2.37 & $-2.69$ & 5\,295$\pm$114 & 3.60$\pm$0.52 & $-2.71\pm$0.24 \\
75091-15989 & 5\,058 & 5\,319 & 2.63 & $-2.67$ & 5\,289$\pm$77 & 2.87$\pm$0.28 & $-3.09\pm$0.11 \\
75293-17425 & 5\,012 & 5\,123 & 1.97 & $-2.87$ & 5\,221$\pm$35 & 1.19$\pm$0.10 & $-2.46\pm$0.16 \\
\hline
   \end{tabular}
   \end{table*}
   
All the selected candidates have been observed with Gaia  and have \teff\ estimates in Gaia DR2 as listed in Table~\ref{table:3}. The different \teff\ estimates from Gaia, SPEEM and n-SSPP are all consistent within 2$\sigma$ and  the mean differences relative to Gaia results are 
\teff$_{\textrm{,nSSPP}}-$\teff$_{\textrm{,Gaia}} = 213 \pm 14$ K
and
\teff$_{\textrm{,SPEEM}}-$\teff$_{\textrm{,Gaia}} = 165 \pm 13$ K.
The differences between the effective temperatures estimated with SPEEM and those derived from the spectroscopic analysis vary between $-211$\,K and $+228$\,K, with an average difference of $49\pm128$ K.
   
We repeated our search for stars in common with SEGUE and it turns out that one of  the candidates observed with ISIS/WHT has already been analysed by SEGUE, providing  \teff$_{\textrm{,SEGUE}}$=5115~K and \metal$_{\textrm{SEGUE}}= -2.96$, while  the  effective temperature listed in the Gaia DR2 catalogue is 5058~K.  
The parameters estimated  with SPEEM are \teff$_{\textrm{,SPEEM}}$= 5112~K and 
\metal$_{\textrm{SPEEM}}= -2.92$, while the spectroscopic analysis indicates slightly lower values (\teff$_{\textrm{,nSSPP}}$=5032~K and \metal$_{\textrm{nSSPP}}= -2.57$) for the star J-PLUS ID 66723-1757.

The differences in metallicities derived from SPEEM and n-SSPP are on average $+0.11 \pm 0.33$, and most of the differences are within $\pm 0.4$ dex.  All the eleven selected candidates present \metal $<-2.0$ and seven stars have  spectroscopic \metal$<-2.5$, representing a success rate of $64^{+21}_{-29}$\%{\footnote{Fractional uncertainties represented by the Wilson score \citep{wilson1927}}}. This result supports SPEEM as a tool to obtain photometric estimates of the fundamental stellar parameters. 
  
\section{Summary and Conclusions}
  
We present the SPEEM pipeline as a tool to provide the star$\times$QSO classification and to obtain the three basic stellar parameters \teff, \logg, and \metal\, based on photometric measurements in the J-PLUS system.
SPEEM employs RF and XGB machine learning algorithms, trained with samples selected from cross-matched data with SEGUE.
The pipeline is able to recover the parameters estimated by SEGUE within a deviation of $41\pm61$ K for \teff; $0.11\pm0.17$ for \logg, and $0.09\pm0.14$ for \metal.
The SPEEM performance test includes a sample of stars in common with LAMOST. Even though the sample J-PLUS$\times$LAMOST is larger than J-PLUS$\times$SEGUE, the metallicity distribution of the former sample is more unbalanced. As a result, we note a trend in the metallicity difference as a function of temperature difference, representing a variation of $\sim$0.2~dex in \metal\ for  $\sim$100~K in \teff, although the obtained \metal\ shows no significant differences relative to the LAMOST \metal\ estimations.
      
The Gold Sample is a subsample of the J-PLUS DR2, selected according to criteria based on the quality of the photometric measurements and star/galaxy classification from PDF analysis.
The application of SPEEM to the Gold Sample produced stellar parameters for 746\,531 stars. A list of 177 stars with \metal$<-2.5$, \teff$<$5\,500 K, and magnitude $g<17.0$ has been selected as potential candidate very metal-poor stars that are bright enough for a spectroscopic follow-up.
      
Eleven candidates were observed with the WHT+ISIS and the resulting spectra have been analysed using the n-SSPP, yielding spectroscopic values of \teff, \logg, and \metal. The comparison between the parameters estimated by SPEEM and those derived spectroscopically shows that all the studied stars are confirmed very metal-poor stars with \metal$<-2.0$, including 7 stars with \metal$<-2.5$, and one new Extremely Metal-Poor star, with \metal$=-3.09$.

SPEEM presented a success rate of $64^{+21}_{-29}$\% in validating the search for stars with \metal$<-2.5$.
For comparison, the spectroscopic follow-up conducted by the Pristine collaboration has a success rate of 70\% for \metal$<-2.5$ \citep{pristine3} in a sample of 149 stars. In addition, the Best \& Brightest survey \citep{schlaufman2014} found $\sim$ 32\% stars with $-3.0<$\metal$<-2.0$ in their spectroscopic follow-up. Future spectroscopic samples of low-metallicity stars in the \metal$<-3.0$ regime will be used to extend the capabilities of SPEEM, which will be applied to future data releases of the J-PLUS survey.

\begin{acknowledgements}
We thank the referee for suggestions and comments that contributed to the improvement of this paper.
Based on observations made with the JAST80 telescope at the Observatorio Astrofísico de Javalambre (OAJ), in  Teruel, owned, managed, and operated by the Centro de  Estudios de Física del Cosmos de Aragón. We acknowledge the OAJ Data Processing and Archiving Unit (UPAD) for reducing  the OAJ data used in this work. Funding for the J-PLUS  Project has been provided by the Governments of Spain and  Aragón through the Fondo de Inversiones de Teruel; the Aragón Government through the Reseach Groups E96, E103, and $E16\_17R$; the Spanish Ministry of Science, Innovation and Universities (MCIU/AEI/FEDER, UE) with grants PGC2018-097585-B-C21 and PGC2018-097585-B-C22; the Spanish Ministry of Economy and Competitiveness (MINECO)  under AYA2015-66211-C2-1-P, AYA2015-66211-C2-2,  AYA2012-30789, and ICTS-2009-14; and European FEDER  funding (FCDD10-4E-867, FCDD13-4E-2685). The Brazilian  agencies FINEP, FAPESP, and the National Observatory of  Brazil have also contributed to this project. The work of V.M.P. is supported by NOIRLab, which is managed by the Association of Universities for Research in Astronomy (AURA) under a  cooperative agreement with the National Science Foundation. Guoshoujing  Telescope (the Large Sky Area Multi-Object Fiber  Spectroscopic Telescope LAMOST) is a National Major  Scientific Project built by the Chinese Academy of Sciences.  Funding for the project has been provided by the National  Development and Reform Commission. LAMOST is operated and managed by the National Astronomical Observatories, Chinese  Academy of Sciences. Y.S.L. acknowledges support from the National Research Foundation (NRF) of Korea grant funded by the Ministry of Science and ICT (NRF-2021R1A2C1008679). 
F.J.E. acknowledges financial support from the Spanish MINECO/FEDER through the grant AYA2017-84089 and MDM-2017-0737 at Centro de Astrobiología (CSIC-INTA), Unidad de Excelencia María de Maeztu, and from the European Union’s Horizon 2020 research and innovation programme under Grant Agreement no. 824064 through the ESCAPE - The European Science Cluster of Astronomy \& Particle Physics ESFRI Research Infrastructures project. R.A.D. acknowledges support from the CNPq through BP grant 308105/2018-4.
This research has made use of the Spanish Virtual Observatory (http://svo.cab.inta-csic.es) supported from the Spanish MICINN/FEDER through grant AyA2017-84089. This research made use of Matplotlib, a 2D graphics package used for Python for publication-quality  image generation across user interfaces and operating  systems (Hunter 2007).
\end{acknowledgements}

\bibliographystyle{aa}
\bibliography{references.bib}

\end{document}